\documentclass{article}
\usepackage{amsmath}

\numberwithin{equation}{section}
\input{tcilatex}
\begin{document}

\title{The SLOCC invariant and the residual entanglement for n-qubits 
\thanks{%
The paper was supported by NSFC(Grant No. 60433050), the basic research fund
of Tsinghua university No: JC2003043 and partially by the state key lab. of
intelligence technology and system}}
\author{Dafa Li \\
Dept of mathematical sciences\\
Tsinghua University, Beijing 100084 CHINA\\
email address: dli@math.tsinghua.edu.cn \and Xiangrong Li \\
Department of Mathematics\\
University of California, Irvine, CA 92697-3875, USA \and Hongtao Huang \\
Electrical Engineering and Computer Science Department\\
University of Michigan, Ann Arbor, MI 48109, USA \and Xinxin Li \\
Dept. of computer science\\
Wayne State University, Detroit, MI 48202, USA}
\maketitle

\begin{abstract}
In this paper, we find the invariant for $n$-qubits and propose the residual
entanglement for $n$-qubits by means of the invariant. Thus, we establish a
relation between SLOCC entanglement and the residual entanglement. The
invariant and the residual entanglement can be used for SLOCC entanglement
classification for $n$-qubits.
\end{abstract}














\section{Introduction}

Entanglement plays a key role in quantum computing and quantum information.
If two states can be obtained from each other by means of local operations
and classical communication (LOCC) with nonzero probability, we say that two
states have the same kind of entanglement\cite{Bennett}. Recently, many
authors have studied the equivalence classes of three-qubit states specified
SLOCC (stochastic local operations and classical communication ) \cite{Acin}$%
-$\cite{LDF}. D\"{u}r et al. showed that for pure states of three-qubits
there are six inequivalent entanglement classes\cite{Dur}. A. Miyake
discussed the onionlike classification of SLOCC\ orbits and proposed the
SLOCC equivalence classes using the orbits\cite{Miyake03}. A.K. Rajagopal
and R.W. Rendell gave the conditions for the full separability and the
biseparability\cite{Rajagopal}. In \cite{LDF-PLA} we gave the simple
criteria for the complete SLOCC\ classification for three-qubits. In \cite%
{LDFJPA} we presented the invariant for 4-qubits and used the invariant for
SLOCC entanglement classification for 4-qubits. Verstraete et al.\cite{Moor2}
considered the entanglement classes of four-qubits under SLOCC and concluded
that there exist nine families of states corresponding to nine different
ways of entanglement.

Coffman et al. presented the concurrence and the residual entanglement for 2
and 3-qubits\cite{Coffman}. It was proven that the residual entanglement for
3-qubits or 3-tangle is an entanglement monotone\cite{Dur}. The general
residual entanglement was discussed in \cite{Yu}. Wong and Nelson presented $%
n$-tangle for even $n$-qubits\cite{Wong}. For odd $n$-qubits, they did not
define $n$-tangle. Osterloh and Siewert constructed $N$-qubit entanglement
monotone from antilinear operators\cite{Osterloh1}\cite{Osterloh2}.

In this paper, we find the SLOCC invariant for $n$-qubits and extend Coffman
et al. 's residual entanglement or 3-tangle for 3-qubits to $n$-qubits in
terms of the invariant. The necessary $D$-criteria and $F$-criteria for
SLOCC classification are also given in this paper. Using the invariant, the
residual entanglement and the criteria, it can be determined that if two
states belong to different SLOCC entanglement classes. The invariant, the
residual entanglement and the criteria only require simple arithmetic
operations: multiplication, addition and subtraction.

The paper is organized as follows. In section 2, we present the invariant
for $n$-qubits and prove the invariant by induction in Appendix D. In
section 3, we propose the residual entanglement for $n$-qubits and
investigate properties of the residual entanglement. In section 4, we
exploit SLOCC entanglement classification for $n$-qubits.

\section{The SLOCC invariant for $n$-qubits}

Let $|\psi \rangle $ and $|\psi ^{\prime }\rangle $\ be any states of $n$%
-qubits. Then we can write

\begin{equation*}
|\psi \rangle =\sum_{i=0}^{2^{n}-1}a_{i}|i\rangle ,|\psi ^{\prime }\rangle
=\sum_{i=0}^{2^{n}-1}b_{i}|i\rangle ,
\end{equation*}

where $\sum_{i=0}^{2^{n}-1}|a_{i}|^{2}=1$ and $%
\sum_{i=0}^{2^{n}-1}|b_{i}|^{2}=1$.

Two states $|\psi \rangle $ and $|\psi ^{\prime }\rangle $ are equivalent
under SLOCC if and only if there exist invertible local operators $\alpha
,\beta $, $\gamma ....$ such that 
\begin{equation}
|\psi \rangle =\underbrace{\alpha \otimes \beta \otimes \gamma \cdots }%
_{n}|\psi ^{\prime }\rangle ,  \label{eqvcond}
\end{equation}

where the local operators $\alpha ,\beta ,\gamma ,....,$ can be expressed as 
$2\times 2$ invertible matrices as follows.

\begin{equation*}
\alpha =\left( 
\begin{array}{cc}
\alpha _{1} & \alpha _{2} \\ 
\alpha _{3} & \alpha _{4}%
\end{array}%
\right) ,~\beta =\left( 
\begin{array}{cc}
\beta _{1} & \beta _{2} \\ 
\beta _{3} & \beta _{4}%
\end{array}%
\right) ,\gamma =\left( 
\begin{tabular}{cc}
$\ \gamma _{1}$ & $\ \gamma _{2}$ \\ 
$\ \gamma _{3}$ & $\ \gamma _{4}$%
\end{tabular}%
\right) ,...
\end{equation*}

We reported the invariants for 2-qubits, 3-qubits and 4-qubits in \cite%
{LDFJPA}. When $n$ is small, by solving the corresponding matrix equations
in (\ref{eqvcond}), we can obtain the amplitudes $a_{i}$. Then, it is easy
to verify the invariants for 2-qubits, 3-qubits and 4-qubits. However, when $%
n$ is large, it is hard to solve the matrix equations in (\ref{eqvcond}).

We define function $sign(n,i)=\pm 1$ to describe the invariant below.

Always $sign(2,0)=sign(3,0)=1$. For $n\geq 4$ and $0\leq i\leq 2^{n-3}-1$,
we define $sign(n,i)$ as follows.



When $0\leq i\leq 2^{n-4}-1$, $sign(n,i)=sign(n-1,i)$. When $2^{n-4}-1<i\leq
2^{n-3}-1$, $sign(n,i)=sign(n,2^{n-3}-1-i)$ provided that $n$ is odd; when $%
n $ is even, $sign(n,i)=-sign(n,2^{n-3}-1-i)$.

\subsection{The SLOCC invariant for even $n$-qubits}

\subsubsection{For 2-qubits}

If $|\psi \rangle $ and $|\psi ^{\prime }\rangle $ are equivalent under
SLOCC, then they satisfy the following equation,

\begin{equation}
a_{0}a_{3}-a_{1}a_{2}=(b_{0}b_{3}-b_{1}b_{2})\det (\alpha )\det (\beta ).
\label{inv21}
\end{equation}%
(\ref{inv21}) guarantees that $(b_{0}b_{3}-b_{1}b_{2})$ does not vary when $%
\det (\alpha )\det (\beta )=1$ or vanish under SLOCC operators $\alpha $ and 
$\beta $.

\subsubsection{For 4-qubits}

$|\psi \rangle $ and $|\psi ^{\prime }\rangle $ are equivalent under SLOCC
if and only if there exist invertible local operators $\alpha $, $\beta $, $%
\gamma $ and $\delta $ such that

\begin{equation}
|\psi \rangle =\alpha \otimes \beta \otimes \gamma \otimes \delta |\psi
^{\prime }\rangle ,  \label{inv41}
\end{equation}

where

\begin{equation*}
\delta =\left( 
\begin{tabular}{cc}
$\ \delta _{1}$ & $\ \delta _{2}$ \\ 
$\ \delta _{3}$ & $\ \delta _{4}$%
\end{tabular}%
\right) .
\end{equation*}

Let 
\begin{equation*}
IV(a,4)=(a_{0}a_{15}-a_{1}a_{14})+(a_{6}a_{9}-a_{7}a_{8})-(a_{2}a_{13}-a_{3}a_{12})-(a_{4}a_{11}-a_{5}a_{10})
\end{equation*}
and 
\begin{equation*}
IV(b,4)=((b_{0}b_{15}-b_{1}b_{14})+(b_{6}b_{9}-b_{7}b_{8})-(b_{2}b_{13}-b_{3}b_{12})-(b_{4}b_{11}-b_{5}b_{10})).
\end{equation*}

Then, if $|\psi \rangle $ and $|\psi ^{\prime }\rangle $ are equivalent
under SLOCC, then we have the following equation: 
\begin{equation}
IV(a,4)=IV(b,4)\ast \det (\alpha )\det (\beta )\det (\gamma )\det (\delta ).
\label{inv42}
\end{equation}

In Appendix A of this paper, we give a formal derivation of (\ref{inv42}).
The ideas for the proof will be used to by induction derive the following
Theorem 1.

By (\ref{inv42}), $IV(b,4)$ does not vary when $\det (\alpha )\det (\beta
)\det (\gamma )\det (\delta )=1$ or vanish under SLOCC operators.

\subsubsection{The definition and proof of the invariant for even $n$-qubits}

Let $|\psi \rangle $ and $|\psi ^{\prime }\rangle $\ be any pure states of $%
n $-qubits.

\textbf{Version 1 of the invariant}

When $n\geq 4$, let 
\begin{eqnarray}
&&IV(a,n)=  \notag \\
&&%
\sum_{i=0}^{2^{n-3}-1}sign(n,i)[(a_{2i}a_{(2^{n}-1)-2i}-a_{2i+1}a_{(2^{n}-2)-2i})
\notag \\
&&+(a_{(2^{n-1}-2)-2i}a_{(2^{n-1}+1)+2i}-a_{(2^{n-1}-1)-2i}a_{2^{n-1}+2i})].
\label{EvenIV1}
\end{eqnarray}
\ Theorem 1.

For $n(\geq 4)$-qubits, assume that $|\psi \rangle $ and $|\psi ^{\prime
}\rangle $ are equivalent under SLOCC. Then the amplitudes of the two states
satisfy the following equation, 
\begin{equation}
IV(a,n)=IV(b,n)\underbrace{\det (\alpha )\det (\beta )\det (\gamma )...}_{n},
\label{EvenIV2}
\end{equation}%
where $IV(b,n)$ is obtained from $IV(a,n)$ by replacing $a$ in $IV(a,n)$\ by 
$b$.

An inductive proof of Theorem 1 is put in Part 1 of Appendix D.

By (\ref{EvenIV2}), clearly $IV(b,n)$ does not vary when $\det (\alpha )\det
(\beta )\det (\gamma )...=1$ or vanish under\textbf{\ }SLOCC operators. So,
here, $IV(b,n)$ is called as an invariant of even $n$-qubits.

So far, no one has reported the invariant for 6-qubits. Therefore, it is
valuable to verify that (\ref{EvenIV2}) holds when $n=6$.

For 6-qubits,

$|\psi \rangle $ and $|\psi ^{\prime }\rangle $ are equivalent under SLOCC
if and only if there exist invertible local operators $\alpha $, $\beta $, $%
\gamma $, $\delta $, $\sigma $\ and $\tau $\ such that

\begin{equation}
|\psi \rangle =\alpha \otimes \beta \otimes \gamma \otimes \delta \otimes
\sigma \otimes \tau |\psi ^{\prime }\rangle ,  \label{inv61}
\end{equation}

where $\sigma =\left( 
\begin{tabular}{cc}
$\ \sigma _{1}$ & $\ \sigma _{2}$ \\ 
$\ \sigma _{3}$ & $\ \sigma _{4}$%
\end{tabular}%
\right) $ and $\tau =\left( 
\begin{tabular}{cc}
$\ \tau _{1}$ & $\ \tau _{2}$ \\ 
$\ \tau _{3}$ & $\ \tau _{4}$%
\end{tabular}%
\right) $.

From (\ref{EvenIV1}),

\begin{eqnarray}
&&IV(a,6)=  \notag \\
&&(a_{0}a_{63}-a_{1}a_{62})+(a_{30}a_{33}-a_{31}a_{32})-(a_{2}a_{61}-a_{3}a_{60})-(a_{28}a_{35}-a_{29}a_{34})
\notag \\
&&-(a_{4}a_{59}-a_{5}a_{58})-(a_{26}a_{37}-a_{27}a_{36})+(a_{6}a_{57}-a_{7}a_{56})+(a_{24}a_{39}-a_{25}a_{38})
\notag \\
&&-(a_{8}a_{55}-a_{9}a_{54})-(a_{22}a_{41}-a_{23}a_{40})+(a_{10}a_{53}-a_{11}a_{52})+(a_{20}a_{43}-a_{21}a_{42})
\notag \\
&&+(a_{12}a_{51}-a_{13}a_{50})+(a_{18}a_{45}-a_{19}a_{44})-(a_{14}a_{49}-a_{15}a_{48})-(a_{16}a_{47}-a_{17}a_{46}).
\notag
\end{eqnarray}

By solving the complicated matrix equation in (\ref{inv61}) by using
MATHEMATICA, we obtain the amplitudes $a_{i}$. Each $a_{i}$ is an algebraic
sum of 64 terms being of the form $b_{j}\alpha _{k}\beta _{l}\gamma
_{m}\delta _{s}\sigma _{t}\tau _{h}$.\ Then, by substituting $a_{i}$ into $%
IV(a,6)$, we obtain the following. 
\begin{equation}
IV(a,6)=IV(b,6)\det (\alpha )\det (\beta )\det (\gamma )\det (\delta )\det
(\sigma )\det (\tau ).  \label{inv62}
\end{equation}

\textbf{Version 2 of the invariant}

Definition

$sign^{\ast }(2,0)=1$. When $n\geq 3$, $sign^{\ast }(n,i)=sign(n,i)$
whenever $0\leq i\leq 2^{n-3}-1$ and $sign^{\ast }(n,i)=sign(n,2^{n-2}-1-i)$
whenever $2^{n-3}-1<i\leq 2^{n-2}-1$. \ 

When $n\geq 2$, let 
\begin{eqnarray}
IV^{\ast }(a,n)=\sum_{i=0}^{2^{n-2}-1}sign^{\ast
}(n,i)(a_{2i}a_{(2^{n}-1)-2i}-a_{2i+1}a_{(2^{n}-2)-2i}).  \label{EvenIV3}
\end{eqnarray}

Clearly, when $n\geq 4$, $IV(a,n)=IV^{\ast }(a,n)$.

Thus, Theorem 1 can be rephrased as follows.

For $n(\geq 2)$-qubits, 
\begin{eqnarray}
IV^{\ast }(a,n)=IV^{\ast }(b,n)\underbrace{\det (\alpha )\det (\beta )\det
(\gamma )...}_{n},  \label{EvenIV4}
\end{eqnarray}

where $IV^{\ast }(b,n)$ is obtained from $IV^{\ast }(a,n)$ by replacing $a$
in $IV^{\ast }(a,n)$\ by $b$. \ \ 

When $n=2$, $4$ and $6$, (\ref{EvenIV4}) is reduced to (\ref{inv21}), (\ref%
{inv42}) and (\ref{inv62}), respectively. $IV^{\ast }(b,n)$ is another
version of the invariant for even $n$-qubits.

\subsection{\ The SLOCC invariant for odd $n$-qubits}

\subsubsection{For 3-qubits}

If $|\psi \rangle $ and $|\psi ^{\prime }\rangle $ are equivalent under
SLOCC, then they satisfy the following equation,

\begin{eqnarray}
&&((a_{0}a_{7}-a_{1}a_{6})-(a_{2}a_{5}-a_{3}a_{4}))^{2}-4(a_{0}a_{3}-a_{1}a_{2})(a_{4}a_{7}-a_{5}a_{6})=
\notag \\
&&\lbrack
((b_{0}b_{7}-b_{1}b_{6})-(b_{2}b_{5}-b_{3}b_{4}))^{2}-4(b_{0}b_{3}-b_{1}b_{2})(b_{4}b_{7}-b_{5}b_{6})]\det^{2}(\alpha )\det^{2}(\beta )\det^{2}(\gamma ).
\label{invari31}
\end{eqnarray}

The above equation can be equivalently replaced by one of the following two
equations. 
\begin{eqnarray}
&&\mbox{(1).}%
((a_{0}a_{7}-a_{3}a_{4})+(a_{1}a_{6}-a_{2}a_{5}))^{2}-4(a_{3}a_{5}-a_{1}a_{7})(a_{2}a_{4}-a_{0}a_{6})=
\notag \\
&&(((b_{0}b_{7}-b_{3}b_{4})+(b_{1}b_{6}-b_{2}b_{5}))^{2}-4(b_{3}b_{5}-b_{1}b_{7})(b_{2}b_{4}-b_{0}b_{6}))\det^{2}(\alpha )\det^{2}(\beta )\det^{2}(\gamma );
\notag \\
&&\mbox{(2).}%
(a_{0}a_{7}-a_{3}a_{4}-(a_{1}a_{6}-a_{2}a_{5}))^{2}-4(a_{1}a_{4}-a_{0}a_{5})(a_{3}a_{6}-a_{2}a_{7})=
\notag \\
&&(b_{0}b_{7}-b_{3}b_{4}-(b_{1}b_{6}-b_{2}b_{5}))^{2}-4(b_{1}b_{4}-b_{0}b_{5})(b_{3}b_{6}-b_{2}b_{7})\det^{2}(\alpha )\det^{2}(\beta )\det^{2}(\gamma ).
\notag
\end{eqnarray}

Let $\overline{IV}(a,3)=(a_{0}a_{7}-a_{1}a_{6})-(a_{2}a_{5}-a_{3}a_{4})$, $%
IV^{\ast }(a,2)=a_{0}a_{3}-a_{1}a_{2}$ and $\ IV_{+4}^{\ast
}(a,2)=(a_{4}a_{7}-a_{5}a_{6})$. Then, (\ref{invari31}) can be rewritten as 
\begin{equation}
(\overline{IV}(a,3))^{2}-4IV^{\ast }(a,2)IV_{+4}^{\ast }(a,2)=[(\overline{IV}%
(b,3))^{2}-4IV^{\ast }(b,2)IV_{+4}^{\ast }(b,2)]\det^{2}(\alpha
)\det^{2}(\beta )\det^{2}(\gamma ),  \label{invari32}
\end{equation}

where $\overline{IV}(b,3)$, $IV^{\ast }(b,2)$ and $IV_{+4}^{\ast }(b,2)$ are
obtained from $\overline{IV}(a,3)$, $IV^{\ast }(a,2)$ and $IV_{+4}^{\ast
}(a,2)$ by replacing $a$ by $b$, respectively.

In Appendix B of this paper, we give a formal proof of (\ref{invari32}). The
ideas for the proof will be used to by induction show the following Theorem
2.

By (\ref{invari32}),$\ (\overline{IV}(b,3))^{2}-4IV^{\ast
}(b,2)IV_{+4}^{\ast }(b,2)$ does not vary when $\det^{2}(\alpha
)\det^{2}(\beta )\det^{2}(\gamma )=1$ or vanish under SLOCC operators.

\subsubsection{For 5-qubits}

So far, no one has reported the invariant for 5-qubits. Therefore, it is
worth listing the explicit expression of the invariant for 5-qubits to
understand the complicated expression of the invariant for\ odd $n$-qubits
which is manifested below.

$|\psi \rangle $ and $|\psi ^{\prime }\rangle $ are equivalent under SLOCC
if and only if there exist invertible local operators $\alpha $, $\beta $, $%
\gamma $, $\delta $ and $\sigma $\ such that

\begin{equation}
|\psi \rangle =\alpha \otimes \beta \otimes \gamma \otimes \delta \otimes
\sigma |\psi ^{\prime }\rangle .  \label{inv51}
\end{equation}

\noindent Let 
\begin{eqnarray}
&&A^{\ast
}=[-(a_{2}a_{29}-a_{3}a_{28}-a_{12}a_{19}+a_{13}a_{18})-(a_{4}a_{27}-a_{5}a_{26}-a_{10}a_{21}+a_{11}a_{20})
\notag \\
&&+(a_{0}a_{31}-a_{1}a_{30}-a_{14}a_{17}+a_{15}a_{16})+(a_{6}a_{25}-a_{7}a_{24}-a_{8}a_{23}+a_{9}a_{22})]^{2}
\notag \\
&&-4[(a_{0}a_{15}-a_{1}a_{14})+(a_{6}a_{9}-a_{7}a_{8})-(a_{2}a_{13}-a_{3}a_{12})-(a_{4}a_{11}-a_{5}a_{10})]
\notag \\
&&[(a_{16}a_{31}-a_{17}a_{30})+(a_{22}a_{25}-a_{23}a_{24})-(a_{18}a_{29}-a_{19}a_{28})-(a_{20}a_{27}-a_{21}a_{26})]
\notag
\end{eqnarray}

and let $B^{\ast }$ be obtained from $A^{\ast }$ by replacing $a$ in $%
A^{\ast }$ by $b$. 

Then if $|\psi \rangle $ and $|\psi ^{\prime }\rangle $ are equivalent under
SLOCC, then the amplitudes of the two states satisfy the following equation, 
\begin{equation}
A^{\ast }=B^{\ast }\ast \det^{2}(\alpha )\det^{2}(\beta )\det^{2}(\gamma
)\det^{2}(\delta )\det^{2}(\sigma ).  \label{inv52}
\end{equation}

We have verified (\ref{inv52}) by using MATHEMATICA. That is, by solving the
complicated matrix equation in (\ref{inv51}), we obtain the amplitudes $%
a_{i} $. Each $a_{i}$ is an algebraic sum of 32 terms being of the form $%
b_{j}\alpha _{k}\beta _{l}\gamma _{m}\delta _{s}\sigma _{t}$. Then, by
substituting $a_{i}$ into $A^{\ast }$, we obtain (\ref{inv52}). However,
this verification is helpless to finding a formal proof of the following
Theorem 2. Hence, it is necessary to give a formal argument of (\ref{inv52})
for readers to readily follow the complicated deduction in Appendix D of the
following Theorem 2. The formal argument of (\ref{inv52}) is put in Appendix
C and gives hints which are used to by induction prove the following Theorem
2.

By (\ref{inv52}), $B^{\ast }$ does not vary when $\det^{2}(\alpha
)\det^{2}(\beta )\det^{2}(\gamma )\det^{2}(\delta )\det^{2}(\sigma )=1$ or
vanish under SLOCC operators.

\subsubsection{The definition and proof of SLOCC\ invariant for odd $n$%
-qubits}

Let $|\psi \rangle $ and $|\psi ^{\prime }\rangle $\ be any pure states of $%
n $($\geq 3$)-qubits. Let 
\begin{eqnarray}
&&\overline{IV}(a,n)=  \notag \\
&&%
\sum_{i=0}^{2^{n-3}-1}sign(n,i)[(a_{2i}a_{(2^{n}-1)-2i}-a_{2i+1}a_{(2^{n}-2)-2i})
\notag \\
&&-(a_{(2^{n-1}-2)-2i}a_{(2^{n-1}+1)+2i}-a_{(2^{n-1}-1)-2i}a_{2^{n-1}+2i})].
\label{OddIV1}
\end{eqnarray}

Let $IV_{+2^{n-1}}^{\ast }(a,n-1)$ be obtained from $IV^{\ast }(a,n-1)$ by
adding $2^{n-1}$ to the subscripts in $IV^{\ast }(a,n-1)$ as follows.

\begin{equation*}
IV_{+2^{n-1}}^{\ast }(a,n-1)=\sum_{i=0}^{2^{n-3}-1}sign^{\ast
}(n-1,i)(a_{2^{n-1}+2i}a_{(2^{n}-1)-2i}-a_{2^{n-1}+1+2i}a_{(2^{n}-2)-2i}).
\end{equation*}

For example, $IV^{\ast }(a,2)=a_{0}a_{3}-a_{1}a_{2}$. Then $IV_{+4}^{\ast
}(a,2)=$ $a_{4}a_{7}-a_{5}a_{6}$. \ 

Theorem 2.\ \ 

Assume that $|\psi \rangle $ and $|\psi ^{\prime }\rangle $ are equivalent
under SLOCC. Then the amplitudes of the two states satisfy the following
equation,

\begin{eqnarray}
&&(\overline{IV}(a,n))^{2}-4IV^{\ast }(a,n-1)IV_{+2^{n-1}}^{\ast }(a,n-1)= 
\notag \\
&&[(\overline{IV}(b,n))^{2}-4IV^{\ast }(b,n-1)IV_{+2^{n-1}}^{\ast }(b,n-1)]%
\underbrace{\det^{2}(\alpha )\det^{2}(\beta )\det^{2}(\gamma )...}_{n},
\label{OddIV3}
\end{eqnarray}

where $IV^{\ast }(b,n-1)$ and $IV_{+2^{n-1}}^{\ast }(b,n-1)$ are obtained
from $IV^{\ast }(a,n-1)$ and $IV_{+2^{n-1}}^{\ast }(a,n-1)$ by replacing $a$
by $b$, respectively.

An inductive proof of Theorem 2 is put in Part 2 of Appendix D. When $n=3$
and $5$, (\ref{OddIV3}) becomes (\ref{invari32}) and (\ref{inv52}),
respectively.

(\ref{OddIV3}) declares that $(\overline{IV}(b,n))^{2}-4IV^{\ast
}(b,n-1)IV_{+2^{n-1}}^{\ast }(b,n-1)$ does not vary when $\det^{2}(\alpha
)\det^{2}(\beta )\det^{2}(\gamma )...=1$ or vanish under\textbf{\ }SLOCC
operators. Here, $(\overline{IV}(b,n))^{2}-4IV^{\ast
}(b,n-1)IV_{+2^{n-1}}^{\ast }(b,n-1)$ is called as an invariant of odd $n$%
-qubits.

\section{The residual entanglement for $n$-qubits}

Coffman et al. \cite{Coffman} defined the residual entanglement for
3-qubits. We propose the residual entanglement for $n$-qubits as follows.

\subsection{The residual entanglement for$\ $even $n$-qubits}

Wong and Nelson's $n$-tangle for even $n$-qubits is listed as follows. See
(2) in \cite{Wong}. 
\begin{equation*}
\tau _{1...n}=2|\sum a_{\alpha _{1}...\alpha _{n}}a_{\beta _{1}...\beta
_{n}}a_{\gamma _{1}...\gamma _{n}}a_{\delta _{1}...\delta _{n}}\times
\epsilon _{\alpha _{1}\beta _{1}}\epsilon _{\alpha _{2}\beta
_{2}}...\epsilon _{\alpha _{n-1}\beta _{n-1}}\epsilon _{\gamma _{1}\delta
_{1}}\epsilon _{\gamma _{2}\delta _{2}}....\times \epsilon _{\gamma
_{n-1}\delta _{n-1}}\epsilon _{\alpha _{n}\gamma _{n}}\epsilon _{\beta
_{n}\delta _{n}}|.
\end{equation*}

The $n$-tangle requires $3\ast 2^{4n}$ multiplications.

When $n$ is even, by means of (\ref{EvenIV3}),\ i.e., the invariant for even 
$n$-qubits, we define that for any state $|\psi \rangle $, the residual
entanglement 
\begin{equation}
\tau (\psi )=2\left\vert IV^{\ast }(a,n)\right\vert .  \label{evenre1}
\end{equation}

This residual entanglement requires $2^{n-1}$ multiplications. When $n=2$,
the residual entanglement $2\left\vert IV^{\ast }(a,2)\right\vert $ just is
Coffman et al. 's concurrence $2\sqrt{\det \rho _{A}}$ \cite{Coffman}.

From Theorem 1, we have the following corollary.

Corollary 1.

If $|\psi \rangle $ and $|\psi ^{\prime }\rangle $ are equivalent under
SLOCC, then from (\ref{EvenIV4}),

\begin{equation}
\tau (\psi )=\tau (\psi ^{\prime })\underbrace{|\det (\alpha )\det (\beta
)\det (\gamma )...|}_{n}.  \label{evenre2}
\end{equation}

It is straightforward to verify the following properties.

Lemma 1.

If a state of even $n$-qubits is a tensor product of a state of 1-qubit and
a state of $(n-1)$-qubits, then $\tau =0$.

In particular, if a state of even $n$-qubits is full separable, then $\tau
=0 $.

Lemma 2.

For 4-qubits, if $|\psi \rangle $ is a tensor product of state $|\phi
\rangle $ of 2-qubits and state $|\omega \rangle $ of 2-qubits, then $\tau
(\psi )=\tau (\phi )\tau (\omega )$.

For 6-qubits, there are two cases.

Case 1. If $|\psi \rangle $ is a tensor product of state $|\phi \rangle $ of
2-qubits and state $|\omega \rangle $ of 4-qubits, then $\tau (\psi )=\tau
(\phi )\tau (\omega )$.

Case 2. If $|\psi \rangle $ is a tensor product of state $|\phi \rangle $ of
3-qubits and state $|\omega \rangle $ of 3-qubits, then $\tau (\psi )=0$.

Conjecture:

(1). If $|\psi \rangle $ is a tensor product of state $|\phi \rangle $ of $%
even$-qubits and state $|\omega \rangle $ of $even$-qubits, then $\tau (\psi
)=\tau (\phi )\tau (\omega )$.

(2). If $|\psi \rangle $ is a tensor product of state $|\phi \rangle $ of $%
odd$-qubits and state $|\omega \rangle $ of $odd$-qubits, then $\tau (\psi
)=0$.

\subsubsection{$\protect\tau \leq 1$}

$|IV^{\ast }(a,n)|\leq \sum_{j=0}^{2^{n-1}-1}\left\vert
a_{j}a_{(2^{n}-1)-j}\right\vert \leq \frac{1}{2}%
\sum_{j=0}^{2^{n-1}-1}(|a_{j}|^{2}+|a_{(2^{n}-1)-j}|^{2})=\frac{1}{2}$.
Therefore $\tau \leq 1$. When $\tau =1$, $\left\vert
a_{j}|=|a_{(2^{n}-1)-j}\right\vert $, where $j=0,1,..,2^{n-1}-1$.

\subsection{The residual entanglement for$\ $odd $n$-qubits}

Wong and Nelson did not discuss odd $n$-tangle\cite{Wong}. When $n$ is odd,
by means of the invariant for odd $n$-qubits, we define that for any state $%
|\psi \rangle $, the residual entanglement 
\begin{equation}
\tau (\psi )=4|(\overline{IV}(a,n))^{2}-4IV^{\ast
}(a,n-1)IV_{+2^{n-1}}^{\ast }(a,n-1)|.  \label{oddre1}
\end{equation}

When $n=3$, this residual entanglement $\tau $ just is Coffman et al. 's
residual entanglement or 3-tangle $\tau _{ABC}=4\left\vert
d_{1}-2d_{2}+4d_{3}\right\vert $\cite{Coffman}.

From Theorem 2, we have the following corollary.

Corollary 2.

If $|\psi \rangle $ and $|\psi ^{\prime }\rangle $ are equivalent under
SLOCC, then by Theorem 2, we obtain

\begin{equation}
\tau (\psi )=\tau (\psi ^{\prime })\underbrace{|\det^{2}(\alpha
)\det^{2}(\beta )\det^{2}(\gamma )...|}_{n}.  \label{oddre2}
\end{equation}

The following results follow the definition of the residual entanglement
immediately.

Lemma 3.

If a state of odd $n$-qubits is a tensor product of a state of 1-qubit and a
state of $(n-1)$-qubits, then $\tau =0$.

In particular, if a state of odd $n$-qubits is full separable, then $\tau =0$%
.

\subsubsection{$\protect\tau \leq 1$}

The fact can be shown by computing the extremes. See Appendix E for the
details. When $\tau =1$, $\left\vert a_{j}|=|a_{(2^{n}-1)-j}\right\vert $,
where $j=0,1,..,2^{n-1}-1$.

\subsection{The invariant residual entanglement}

Corollaries 1 and 2 imply that the residual entanglement does not vary when $%
\left\vert \det (\alpha )\det (\beta )\det (\gamma )..\right\vert =1$ or
vanish under\textbf{\ }SLOCC\ operators. Also, from Corollaries 1 and 2, it
is easy to see that if $|\psi \rangle $ and $|\psi ^{\prime }\rangle $ are
equivalent under SLOCC, then either $\tau (\psi )=\tau (\psi ^{\prime })=0$
or $\tau (\psi )\tau (\psi ^{\prime })\neq 0$. \ Otherwise, the two states
belong to different SLOCC classes.

\subsection{States with the maximal residual entanglement}

(1). Let state $|GHZ\rangle $ of $n$-qubits be $(|\underbrace{0...0}%
_{n}\rangle +|\underbrace{1...1}_{n}\rangle )/\sqrt{2}$. Then, no matter how 
$n$ is even or odd, it is easy to see that $\tau =1$ for state $|GHZ\rangle $
of $n$-qubits. We have shown that $\tau \leq 1$.\ Therefore,\ state $%
|GHZ\rangle $ has the maximal residual entanglement, i.e., $\tau =1$. Also, $%
\tau =1$ for any state of $n$-qubits which is equivalent to $|GHZ\rangle $
under determinant one SLOCC\ operations.

(2). There are many true entangled states with the maximal residual
entanglement.

For example, when $n=4$, $|C\rangle =(|3\rangle +|5\rangle +|6\rangle
+|9\rangle +|10\rangle +|12)/\sqrt{6}$ \cite{LDF-PLA}. $\tau (C)=1$. As
well, $\tau =1$ for any state of $4$-qubits which is equivalent to $%
|C\rangle $ under determinant one SLOCC\ operations. \ \ 

(3) There are many product states\ with the maximal residual entanglement.

When $n=4$, $\tau =1$ for any state which is equivalent to $|GHZ\rangle
_{12}\otimes |GHZ\rangle _{34}$, $|GHZ\rangle _{13}\otimes |GHZ\rangle _{24}$
or $|GHZ\rangle _{14}\otimes |GHZ\rangle _{23}$ under determinant one SLOCC\
operations.

When $n=6$, $|GHZ\rangle _{12}\otimes |GHZ\rangle _{3456}$ and $|GHZ\rangle
_{12}\otimes |GHZ\rangle _{34}\otimes |GHZ\rangle _{56\text{ }}$have the
maximal residual entanglement $\tau =1$.

The examples above illustrate that the residual entanglement is not the $n$%
-way entanglement.

\subsection{The true entanglement classes with the minimal residual
entanglement}

(1). For state $|W\rangle $ of $n$-qubits, no matter how $n$ is even($\geq 4$%
) or odd($\geq 3$), $\tau (W)=0$. By Corollaries 1 and 2, $\tau =0$ for any
state which is equivalent to $|W\rangle $ under SLOCC.\ 

(2). For 4-qubits, there are many true SLOCC entanglement classes which have
the minimal residual entanglement $\tau =0$\cite{LDF-PLA}.

\section{SLOCC classification}

We used the invariant, $D$-criteria and $F$-criteria for SLOCC
classification of 4-qubits\cite{LDFJPA}. The invariant and residual
entanglement for $n$-qubits and the following $D$-criteria and $F$-criteria
for $n$-qubits can be used for SLOCC classification of $n$-qubits. In this
section, we also show that the dual states are SLOCC equivalent.

\subsection{$D-$ criteria for $n\geq 4$-qubits}

$%
D_{1}^{(i)}=(a_{1+8i}a_{4+8i}-a_{0+8i}a_{5+8i})(a_{2^{n}-8i-5}a_{2^{n}-8i-2}-a_{2^{n}-8i-6}a_{2^{n}-8i-1}) 
$

$%
-(a_{3+8i}a_{6+8i}-a_{2+8i}a_{7+8i})(a_{2^{n}-8i-7}a_{2^{n}-8i-4}-a_{2^{n}-8i-8}a_{2^{n}-8i-3}), 
$

$%
D_{2}^{(i)}=(a_{4+8i}a_{7+8i}-a_{5+8i}a_{6+8i})(a_{2^{n}-8i-8}a_{2^{n}-8i-5}-a_{2^{n}-8i-7}a_{2^{n}-8i-6}) 
$

$%
-(a_{0+8i}a_{3+8i}-a_{1+8i}a_{2+8i})(a_{2^{n}-8i-4}a_{2^{n}-8i-1}-a_{2^{n}-8i-3}a_{2^{n}-8i-2}), 
$

$%
D_{3}^{(i)}=(a_{3+8i}a_{5+8i}-a_{1+8i}a_{7+8i})(a_{2^{n}-8i-6}a_{2^{n}-8i-4}-a_{2^{n}-8i-8}a_{2^{n}-8i-2}) 
$

$%
-(a_{2+8i}a_{4+8i}-a_{0+8i}a_{6+8i})(a_{2^{n}-8i-5}a_{2^{n}-8i-3}-a_{2^{n}-8i-7}a_{2^{n}-8i-1}) 
$

$i=0,1,...,2^{n-4}-1$.

\subsection{$F-$criteria}

When $i+j$ is odd,

$%
(a_{i}a_{j}+a_{k}a_{l}-a_{p}a_{q}-a_{r}a_{s})^{2}-4(a_{i}a_{j-1}-a_{p}a_{q-1})(a_{k}a_{l+1}-a_{r}a_{s+1}) 
$.

Otherwise,

$%
(a_{i}a_{j}+a_{k}a_{l}-a_{p}a_{q}-a_{r}a_{s})^{2}-4(a_{i}a_{j-2}-a_{p}a_{q-2})(a_{k}a_{l+2}-a_{r}a_{s+2}) 
$.

The subscripts above satisfy the following conditions. 
\begin{eqnarray}
&&i<j,k<l,p<q,r<s,i<k<p<r  \notag \\
&&i+j=k+l=p+q=r+s,i\oplus j=k\oplus l=p\oplus q=r\oplus s.  \label{concur}
\end{eqnarray}

For example, $F$-criteria include expressions in which $i+j=7,11,13,15,17,19$
and $23$ and the expressions in which $i+j=14$ and $16,$ exclude the
expressions in which $i+j=8,9,10,12,18,20,21$ or $22$.

\subsection{The dual states are SLOCC equivalent}

Let $\bar{1}$ ( $\bar{0}$ ) be the complement of a bit 1 $(0)$. Then $\bar{0}
$ $=1$ and $\bar{1}=0$. Let $\bar{z}=\bar{z_{1}}\bar{z_{2}}...\bar{z_{n}}$
denote the complement of a binary string $z=z_{1}z_{2}....z_{n}$. Also, the
set of the basis states $B=\{|\bar{0}\rangle ,|\bar{1}\rangle ,...,|%
\overline{2^{n}-1}\rangle \}$. Let $|\varphi \rangle $ be any state of $n$%
-qubits. Then we can write $|\varphi \rangle =$ $c_{0}|0\rangle $ $%
+c_{1}|1\rangle $ $+....+c_{2^{n}-1}|(2^{n}-1)\rangle $. Let $|\overline{%
\varphi }\rangle =c_{0}|\bar{0}\rangle $ $+c_{1}|\bar{1}\rangle $ $%
+....+c_{2^{n}-1}|\overline{(2^{n}-1)}\rangle $. We call $|\overline{\varphi 
}\rangle $ the complement of $|\varphi \rangle $.

Let $\sigma _{x}=\left( 
\begin{array}{cc}
0 & 1 \\ 
1 & 0%
\end{array}%
\right) $. Then $\sigma _{x}\otimes ...\otimes \sigma _{x}|\varphi \rangle
=\sum_{i=0}^{2^{n}-1}c_{i}(\sigma _{x}\otimes ...\otimes \sigma
_{x}|i\rangle )=\sum_{i=0}^{2^{n}-1}$ $c_{i}|\bar{\imath}\rangle =|\overline{%
\varphi }\rangle $.

Consequently, if two states of $n$-qubits are dual then they are SLOCC
equivalent.

\section{Summary}

In this paper, we report the invariant for $n$-qubits. The invariant is only
related to the amplitudes of the related two states and the determinants of
the related operators. It reveals the inherent properties of SLOCC
equivalence. By means of the invariant we propose the residual entanglement
for $n$-qubits. When $n=2$, it becomes Coffman et al.'s concurrence for $2$%
-qubits and when $n=3$, it is 3-tangle. For even $n$-qubits, it is much
simpler than Wong and Nelson's even $n$-tangle\cite{Wong}. For odd $n$%
-qubits,\ it requires $2^{n}$ multiplications. Wong and Nelson did not
define the odd $n$-tangle. The properties of the residual entanglement are
discussed in this paper. Wong and Nelson indicated out that when $n$ is
even, $n$-qubit $|GHZ\rangle $ state has the maximal $n$-$\tan $gle and $n$%
-qubit $|W\rangle $ state has the minimal $n$-$\tan $gle\cite{Wong}. The
present paper gives many true entangled states with the maximal residual
entanglement: $\tau =1$ and many true SLOCC entanglement classes with the
minimal residual entanglement: $\tau =0$. Wong and Nelson indicated out that
their even $n$-tangle is not the $n$-way entanglement\cite{Wong}. In the
present paper, the properties of the residual entanglement claim that no
matter how $n$ is even or odd,\ the residual entanglement is not the $n$-way
entanglement. The invariant and the residual entanglement can be used for
SLOCC entanglement classification for $n$-qubits.

\section*{Appendix A: The proof of the invariant for 4-qubits}

\setcounter{equation}{0} \renewcommand{\theequation}{A\arabic{equation}}

Let us prove (\ref{inv42}). We can rewrite 
\begin{equation*}
|\Psi \rangle =|0\rangle \otimes \sum_{i=0}^{7}(\alpha _{1}d_{i}+\alpha
_{2}d_{8+i})|i\rangle +|1\rangle \otimes \sum_{i=0}^{7}(\alpha
_{3}d_{i}+\alpha _{4}d_{8+i})|i\rangle ,
\end{equation*}

where 
\begin{eqnarray}
&&a_{i}=\alpha _{1}d_{i}+\alpha _{2}d_{8+i}\quad \mbox{and}\quad
a_{8+i}=\alpha _{3}d_{i}+\alpha _{4}d_{8+i}, 0\leq i\leq 7,  \label{C0} \\
&&\sum_{i=0}^{7}d_{i}|i\rangle =\beta \otimes \gamma \otimes \delta
\sum_{i=0}^{7}b_{i}|i\rangle,  \label{C1} \\
&& \sum_{i=0}^{7}d_{8+i}|i\rangle =\beta \otimes \gamma \otimes \delta
\sum_{i=0}^{7}b_{8+i}|i\rangle.  \label{C2}
\end{eqnarray}

Notice that from (\ref{C1}) and (\ref{C2}) it happens that $%
\sum_{i=0}^{15}d_{i}|i\rangle =I\otimes \beta \otimes \gamma \otimes \delta
\sum_{i=0}^{15}b_{i}|i\rangle $, where $I$ is an identity.

(\ref{inv42}) follows the following Steps 1 and 2 obviously.

Step 1. Prove $IV(a,4)=IV(d,4)\det (\alpha )$, where $IV(d,4)$ is obtained
from $IV(a,4)$ by replacing $a$ by $d$.

From (\ref{C0}), by computing, 
\begin{equation*}
(a_{2}a_{13}-a_{3}a_{12})+(a_{4}a_{11}-a_{5}a_{10})=[(d_{2}d_{13}-d_{3}d_{12})+(d_{4}d_{11}-d_{5}d_{10})]\det (\alpha ),
\end{equation*}

\begin{equation*}
(a_{0}a_{15}-a_{1}a_{14})+(a_{6}a_{9}-a_{7}a_{8})=[(d_{0}d_{15}-d_{1}d_{14})+(d_{6}d_{9}-d_{7}d_{8})]\det (\alpha ) .
\end{equation*}

So the proof of Step 1 is done.

Step 2. Prove that 
\begin{equation*}
IV(d,4)=IV(b,4)\det (\beta )\det (\gamma )\det (\delta ).
\end{equation*}

We can rewrite (\ref{C1}) as 
\begin{eqnarray}
\sum_{i=0}^{7}d_{i}|i\rangle =|0\rangle \otimes \sum_{i=0}^{3}(\beta
_{1}h_{i}+\beta _{2}h_{4+i})|i\rangle +|1\rangle \otimes
\sum_{i=0}^{3}(\beta _{3}h_{i}+\beta _{4}h_{4+i})|i\rangle,  \label{C4}
\end{eqnarray}

where 
\begin{eqnarray}
&&\sum_{i=0}^{3}h_{i}|i\rangle =\gamma \otimes \delta
\sum_{i=0}^{3}b_{i}|i\rangle,  \label{C5} \\
&& \sum_{i=0}^{3}h_{4+i}|i\rangle =\gamma \otimes \delta
\sum_{i=0}^{3}b_{4+i}|i\rangle,  \label{C6} \\
&&d_{i}=\beta _{1}h_{i}+\beta _{2}h_{4+i}\quad \mbox{and}\quad d_{4+i}=\beta
_{3}h_{i}+\beta _{4}h_{4+i}, 0\leq i\leq 3.  \label{C7}
\end{eqnarray}

Similarly, (\ref{C2}) can be rewritten as 
\begin{equation*}
\sum_{i=0}^{7}d_{8+i}|i\rangle =|0\rangle \otimes \sum_{i=0}^{3}(\beta
_{1}h_{8+i}+\beta _{2}h_{12+i})|i\rangle +|1\rangle \otimes
\sum_{i=0}^{3}(\beta _{3}h_{8+i}+\beta _{4}h_{12+i})|i\rangle ,
\end{equation*}%
where 
\begin{eqnarray}
&&\sum_{i=0}^{3}h_{8+i}|i\rangle =\gamma \otimes \delta
\sum_{i=0}^{3}b_{8+i}|i\rangle ,  \label{C8} \\
&&\sum_{i=0}^{3}h_{12+i}|i\rangle =\gamma \otimes \delta
\sum_{i=0}^{3}b_{12+i}|i\rangle ,  \label{C9} \\
&&d_{8+i}=\beta _{1}h_{8+i}+\beta _{2}h_{12+i}\quad \mbox{and}\quad
d_{12+i}=\beta _{3}h_{8+i}+\beta _{4}h_{12+i},0\leq i\leq 3.  \label{C10}
\end{eqnarray}%
By substituting (\ref{C7}) and (\ref{C10}) into $IV(d,4)$, 
\begin{equation}
IV(d,4)=IV(h,4)\det (\beta ),  \label{C11}
\end{equation}%
where $IV(h,4)$ is obtained from $IV(a,4)$ by replacing $a$ by $h$.

From (\ref{C5}) and (\ref{C6}), 
\begin{eqnarray}
\sum_{i=0}^{7}h_{i}|i\rangle =I\otimes \gamma \otimes \delta
\sum_{i=0}^{7}b_{i}|i\rangle.  \label{C12}
\end{eqnarray}
From (\ref{C8}) and (\ref{C9}), 
\begin{eqnarray}
\sum_{i=0}^{7}h_{8+i}|i\rangle =I\otimes \gamma \otimes \delta
\sum_{i=0}^{7}b_{8+i}|i\rangle .  \label{C13}
\end{eqnarray}
From (\ref{C12}) and (\ref{C13}), 
\begin{eqnarray}
\sum_{i=0}^{15}h_{i}|i\rangle =I\otimes I\otimes \gamma \otimes \delta
\sum_{i=0}^{15}b_{i}|i\rangle .  \label{C14}
\end{eqnarray}

Similarly, from (\ref{C14}) we\ can derive 
\begin{equation}
IV(h,4)=IV(b,4)\det (\gamma )\det (\delta ).  \label{C15}
\end{equation}%
From (\ref{C11}) and (\ref{C15}), the proof of Step 2 is done.

\section*{Appendix B: The proof of the invariant for 3-qubits}

\setcounter{equation}{0} \renewcommand{\theequation}{B\arabic{equation}}

We can rewrite 
\begin{equation*}
|\Psi \rangle =|0\rangle \otimes \sum_{i=0}^{3}(\alpha _{1}d_{i}+\alpha
_{2}d_{4+i})|i\rangle +|1\rangle \otimes \sum_{i=0}^{3}(\alpha
_{3}d_{i}+\alpha _{4}d_{4+i})|i\rangle ,
\end{equation*}%
where 
\begin{eqnarray}
&&a_{i}=\alpha _{1}d_{i}+\alpha _{2}d_{4+i}\quad \mbox{and}\quad
a_{4+i}=\alpha _{3}d_{i}+\alpha _{4}d_{4+i},0\leq i\leq 3,  \label{B0} \\
&&\sum_{i=0}^{3}d_{i}|i\rangle =\beta \otimes \gamma
\sum_{i=0}^{3}b_{i}|i\rangle ,  \label{B1} \\
&&\sum_{i=0}^{3}d_{4+i}|i\rangle =\beta \otimes \gamma
\sum_{i=0}^{3}b_{4+i}|i\rangle .  \label{B2}
\end{eqnarray}

Notice that from (\ref{B1}) and (\ref{B2}) it happens that $%
\sum_{i=0}^{7}d_{i}|i\rangle =I\otimes \beta \otimes \gamma
\sum_{i=0}^{7}b_{i}|i\rangle $, where $I$ is an identity.

(\ref{invari32}) can be obtained from the following Steps 1 and 2.

Step 1. Prove that 
\begin{equation*}
(\overline{IV}(a,3))^{2}-4IV^{\ast }(a,2)IV_{+4}^{\ast }(a,2)=[(\overline{IV}%
(d,3))^{2}-4IV^{\ast }(d,2)IV_{+4}^{\ast }(d,2)]\det^{2}(\alpha ),
\end{equation*}

where $\overline{IV}(d,3)$, $IV^{\ast }(d,2)$ and $IV_{+4}^{\ast }(d,2)$ are
obtained from $\overline{IV}(a,3)$, $IV^{\ast }(a,2)$ and $IV_{+4}^{\ast
}(a,2)$ by replacing $a$ by $d$, respectively.

From (\ref{B0}), by computing, 
\begin{eqnarray}
&&IV^{\ast }(a,2)=IV^{\ast }(d,2)\alpha _{1}^{2}+\overline{IV}(d,3)\alpha
_{1}\alpha _{2}+IV_{+4}^{\ast }(d,2)\alpha _{2}^{2},  \label{B3} \\
&&IV_{+4}^{\ast }(a,2)=IV^{\ast }(d,2)\alpha _{3}^{2}+\overline{IV}%
(d,3)\alpha _{3}\alpha _{4}+IV_{+4}^{\ast }(d,2)\alpha _{4}^{2},  \label{B4}
\\
&&\overline{IV}(a,3)=2IV^{\ast }(d,2)\alpha _{1}\alpha _{3}+\overline{IV}%
(d,3)(\alpha _{1}\alpha _{4}+\alpha _{2}\alpha _{3})+2IV_{+4}^{\ast
}(d,2)\alpha _{2}\alpha _{4}.  \label{B5}
\end{eqnarray}%
Then the proof of Step 1 follows (\ref{B3}), (\ref{B4}) and (\ref{B5})
straightforwardly.

Step 2. Prove that 
\begin{equation*}
(\overline{IV}(d,3))^{2}-4IV^{\ast }(d,2)IV_{+4}^{\ast }(d,2)=[(\overline{IV}%
(b,3))^{2}-4IV^{\ast }(b,2)IV_{+4}^{\ast }(b,2)]\det^{2}(\beta
)\det^{2}(\gamma ).
\end{equation*}%
By (\ref{inv21}), from (\ref{B1}), 
\begin{equation}
IV^{\ast }(d,2)=IV^{\ast }(b,2)\det (\beta )\det (\gamma ),  \label{B6}
\end{equation}

and from (\ref{B2}), 
\begin{equation}
IV_{+4}^{\ast }(d,2)=IV_{+4}^{\ast }(b,2)\det (\beta )\det (\gamma ).
\label{B7}
\end{equation}

Let us compute $\overline{IV}(d,3)$. From (\ref{B1}) and (\ref{B2}) we
obtain 
\begin{eqnarray}
\sum_{i=0}^{3}(d_{i}-d_{4+i})|i\rangle =\beta \otimes \gamma
\sum_{i=0}^{3}(b_{i}-b_{4+i})|i\rangle .  \label{B8}
\end{eqnarray}

By (\ref{inv21}), from (\ref{B8}) it is easy to see 
\begin{equation}
(d_{0}-d_{4})(d_{3}-d_{7})-(d_{1}-d_{5})(d_{2}-d_{6})=[(b_{0}-b_{4})(b_{3}-b_{7})-(b_{1}-b_{5})(b_{2}-b_{6})]\det (\beta )\det (\gamma ).
\label{B9}
\end{equation}

Expanding (\ref{B9}), we have 
\begin{eqnarray}
IV^{\ast }(d,2)+IV_{+4}^{\ast }(d,2)-\overline{IV}(d,3)=[IV^{\ast
}(b,2)+IV_{+4}^{\ast }(b,2)-\overline{IV}(b,3)]\det (\beta )\det (\gamma ).
\label{B10}
\end{eqnarray}
From (\ref{B6}), (\ref{B7}) and (\ref{B10}), we get 
\begin{eqnarray}
\overline{IV}(d,3)=\overline{IV}(b,3)\det (\beta )\det (\gamma ).
\label{B11}
\end{eqnarray}

The proof of Step 2 follows (\ref{B6}), (\ref{B7}) and (\ref{B11})
immediately.

\section*{Appendix C: The proof of the invariant for 5-qubits}

\setcounter{equation}{0} \renewcommand{\theequation}{C\arabic{equation}}

$|\psi ^{\prime }\rangle $ can be rewritten as 
\begin{equation*}
|\psi ^{\prime }\rangle =|0\rangle \otimes \sum_{i=0}^{15}b_{i}|i\rangle
+|1\rangle \otimes \sum_{i=0}^{15}b_{16+i}|i\rangle .
\end{equation*}

Thus, 
\begin{equation*}
|\psi \rangle =\alpha |0\rangle \otimes \beta \otimes \gamma \otimes \delta
\otimes \sigma \sum_{i=0}^{15}b_{i}|i\rangle +\alpha |1\rangle \otimes \beta
\otimes \gamma \otimes \delta \otimes \sigma
\sum_{i=0}^{15}b_{16+i}|i\rangle.
\end{equation*}

Let 
\begin{eqnarray}
\sum_{i=0}^{15}d_{i}|i\rangle =\beta \otimes \gamma \otimes \delta \otimes
\sigma \sum_{i=0}^{15}b_{i}|i\rangle  \label{A1}
\end{eqnarray}
and 
\begin{eqnarray}
\sum_{i=0}^{15}d_{16+i}|i\rangle =\beta \otimes \gamma \otimes \delta
\otimes \sigma \sum_{i=0}^{15}b_{16+i}|i\rangle.  \label{A2}
\end{eqnarray}

By (\ref{A1}) and (\ref{A2}), we can rewrite 
\begin{eqnarray}
|\psi \rangle =(\alpha _{1}|0\rangle +\alpha _{3}|1\rangle )\otimes
\sum_{i=0}^{15}d_{i}|i\rangle +(\alpha _{2}|0\rangle +\alpha _{4}|1\rangle
)\otimes \sum_{i=0}^{15}d_{16+i}|i\rangle.  \label{A3}
\end{eqnarray}

From (\ref{A3}), we have 
\begin{eqnarray}
|\psi \rangle =|0\rangle \otimes \sum_{i=0}^{15}(\alpha _{1}d_{i}+\alpha
_{2}d_{16+i})|i\rangle +|1\rangle \otimes \sum_{i=0}^{15}(\alpha
_{3}d_{i}+\alpha _{4}d_{16+i})|i\rangle.  \label{A4}
\end{eqnarray}

From (\ref{A4}), we can obtain the amplitudes 
\begin{eqnarray}
a_{i}=\alpha _{1}d_{i}+\alpha _{2}d_{16+i}\quad\mbox{and} \quad
a_{16+i}=\alpha _{3}d_{i}+\alpha _{4}d_{16+i},  \label{A5}
\end{eqnarray}

where $0\leq i\leq 15$.

By substituting (\ref{A5}) into $A^{\ast }$, we obtain 
\begin{equation*}
A^{\ast }=D^{\ast }\ast \det^{2}(\alpha ),
\end{equation*}

where 
\begin{eqnarray}
&&D^*=
\{[(d_{2}d_{29}-d_{3}d_{28}-d_{12}d_{19}+d_{13}d_{18})+(d_{4}d_{27}-d_{5}d_{26}-d_{10}d_{21}+d_{11}d_{20})
\notag \\
&&-(d_{0}d_{31}-d_{1}d_{30}-d_{14}d_{17}+d_{15}d_{16})-(d_{6}d_{25}-d_{7}d_{24}-d_{8}d_{23}+d_{9}d_{22})]^{2}
\notag \\
&&-4[(d_{0}d_{15}-d_{1}d_{14})+(d_{6}d_{9}-d_{7}d_{8})-(d_{2}d_{13}-d_{3}d_{12})-(d_{4}d_{11}-d_{5}d_{10})]
\notag \\
&&[(d_{16}d_{31}-d_{17}d_{30})+(d_{22}d_{25}-d_{23}d_{24})-(d_{18}d_{29}-d_{19}d_{28})-(d_{20}d_{27}-d_{21}d_{26})]\} .
\notag
\end{eqnarray}

Next let us show that 
\begin{eqnarray}
D^*=B^*\ast\det^{2}(\beta )\det^{2}(\gamma )\det^{2}(\delta )\det^{2}(\sigma
).  \label{fiveq1}
\end{eqnarray}

From (\ref{A1}) and by (\ref{inv42}), we obtain

\begin{eqnarray}
&&(d_{0}d_{15}-d_{1}d_{14})+(d_{6}d_{9}-d_{7}d_{8})-(d_{2}d_{13}-d_{3}d_{12})-(d_{4}d_{11}-d_{5}d_{10})=
\notag \\
&&[(b_{0}b_{15}-b_{1}b_{14})+(b_{6}b_{9}-b_{7}b_{8})-(b_{2}b_{13}-b_{3}b_{12})-(b_{4}b_{11}-b_{5}b_{10})]
\notag \\
&& \ast \det (\beta )\det (\gamma )\det (\delta )\det (\sigma ).  \label{A7}
\end{eqnarray}

From (\ref{A2}) and by (\ref{inv42}), we obtain

\begin{eqnarray}
&&(d_{16}d_{31}-d_{17}d_{30})+(d_{22}d_{25}-d_{23}d_{24})-(d_{18}d_{29}-d_{19}d_{28})-(d_{20}d_{27}-d_{21}d_{26})=
\notag \\
&&[(b_{16}b_{31}-b_{17}b_{30})+(b_{22}b_{25}-b_{23}b_{24})-(b_{18}b_{29}-b_{19}b_{28})-(b_{20}b_{27}-b_{21}b_{26})]
\notag \\
&&\ast \det (\beta )\det (\gamma )\det (\delta )\det (\sigma ).  \label{A8}
\end{eqnarray}

From (\ref{A1}) and (\ref{A2}), we have 
\begin{eqnarray}
&&\sum_{i=0}^{15}(d_{i}-d_{16+i})|i\rangle =\beta \otimes \gamma \otimes
\delta \otimes \sigma \sum_{i=0}^{15}(b_{i}-b_{16+i})|i\rangle.  \label{A9}
\end{eqnarray}

By (\ref{inv42}), from (\ref{A9}) we obtain

\begin{eqnarray}
&&((d_{0}-d_{16})(d_{15}-d_{31})-(d_{1}-d_{17})(d_{14}-d_{30}))+((d_{6}-d_{22})(d_{9}-d_{25})-(d_{7}-d_{23})(d_{8}-d_{24}))
\notag \\
&&-((d_{2}-d_{18})(d_{13}-d_{29})-(d_{3}-d_{19})(d_{12}-d_{28}))-((d_{4}-d_{20})(d_{11}-d_{27})-(d_{5}-d_{21})(d_{10}-d_{26}))=
\notag \\
&&((b_{0}-b_{16})(b_{15}-b_{31})-(b_{1}-b_{17})(b_{14}-b_{30}))+((b_{6}-b_{22})(b_{9}-b_{25})-(b_{7}-b_{23})(b_{8}-b_{24}))
\notag \\
&&-((b_{2}-b_{18})(b_{13}-b_{29})-(b_{3}-b_{19})(b_{12}-b_{28}))-((b_{4}-b_{20})(b_{11}-b_{27})-(b_{5}-b_{21})(b_{10}-b_{26}))
\notag \\
&&\ast \det (\beta )\det (\gamma )\det (\delta )\det (\sigma ).  \label{A10}
\end{eqnarray}

By expanding (\ref{A10}) and using (\ref{A7}) and (\ref{A8}), we obtain

\begin{eqnarray}
&&(d_{2}d_{29}-d_{3}d_{28}-d_{12}d_{19}+d_{13}d_{18})+(d_{4}d_{27}-d_{5}d_{26}-d_{10}d_{21}+d_{11}d_{20})
\notag \\
&&-(d_{0}d_{31}-d_{1}d_{30}-d_{14}d_{17}+d_{15}d_{16})-(d_{6}d_{25}-d_{7}d_{24}-d_{8}d_{23}+d_{9}d_{22})=
\notag \\
&&[(b_{2}b_{29}-b_{3}b_{28}-b_{12}b_{19}+b_{13}b_{18})+(b_{4}b_{27}-b_{5}b_{26}-b_{10}b_{21}+b_{11}b_{20})
\notag \\
&&-(b_{0}b_{31}-b_{1}b_{30}-b_{14}b_{17}+b_{15}b_{16})-(b_{6}b_{25}-b_{7}b_{24}-b_{8}b_{23}+b_{9}b_{22})]
\notag \\
&&\det (\beta )\det (\gamma )\det (\delta )\det (\sigma ).  \label{A11}
\end{eqnarray}

Then (\ref{fiveq1}) follows (\ref{A7}), (\ref{A8}) and (\ref{A11}).

Finally, (\ref{inv52}) follows (\ref{A6}) and (\ref{fiveq1}).

\section*{Appendix D: The proofs of the invariant for $n$-qubits}

\setcounter{equation}{0} \renewcommand{\theequation}{D\arabic{equation}}

We can rewrite 
\begin{equation*}
|\Psi \rangle =|0\rangle \otimes \sum_{i=0}^{2^{n-1}-1}(\alpha
_{1}d_{i}+\alpha _{2}d_{2^{n-1}+i})|i\rangle +|1\rangle \otimes
\sum_{i=0}^{2^{n-1}-1}(\alpha _{3}d_{i}+\alpha _{4}d_{2^{n-1}+i})|i\rangle ,
\end{equation*}%
where 
\begin{eqnarray}
&&a_{i}=\alpha _{1}d_{i}+\alpha _{2}d_{2^{n-1}+i}\quad \mbox{and}\quad
a_{2^{n-1}+i}=\alpha _{3}d_{i}+\alpha _{4}d_{2^{n-1}+i},  \label{D0} \\
&&0\leq i\leq 2^{n-1}-1,  \notag
\end{eqnarray}%
\begin{eqnarray}
&&\sum_{i=0}^{2^{n-1}-1}d_{i}|i\rangle =\underbrace{\beta \otimes \gamma
\otimes ....}_{n-1}\sum_{i=0}^{2^{n-1}-1}b_{i}|i\rangle ,  \label{D1} \\
&&\quad \sum_{i=0}^{2^{n-1}-1}d_{2^{n-1}+i}|i\rangle =\underbrace{\beta
\otimes \gamma \otimes ....}_{n-1}\sum_{i=0}^{2^{n-1}-1}b_{2^{n-1}+i}|i%
\rangle .  \label{D2}
\end{eqnarray}

From (\ref{D1}) and (\ref{D2}), it happens that $\sum_{i=0}^{2^{n}-1}d_{i}|i%
\rangle =I\otimes \beta \otimes \gamma
.....\sum_{i=0}^{2^{n}-1}b_{i}|i\rangle $, where $I$ is an identity.

Lemma 1.

\begin{eqnarray}
&&(a_{2i}a_{(2^{n}-1)-2i}-a_{2i+1}a_{(2^{n}-2)-2i})+(a_{(2^{n-1}-2)-2i}a_{(2^{n-1}+1)+2i}-a_{(2^{n-1}-1)-2i}a_{2^{n-1}+2i})=
\notag \\
&&(d_{2i}d_{(2^{n}-1)-2i}-d_{2i+1}d_{(2^{n}-2)-2i})+(d_{(2^{n-1}-2)-2i}d_{(2^{n-1}+1)+2i}-d_{(2^{n-1}-1)-2i}d_{2^{n-1}+2i})
\notag \\
&&*\det(\alpha )  \label{lemma1}
\end{eqnarray}

Proof.

By (\ref{D0}), 
\begin{eqnarray}
&& a_{2i}=\alpha _{1}d_{2i}+\alpha _{2}d_{2^{n-1}+2i},  \notag \\
&&a_{(2^{n}-1)-2i}=\alpha _{3}d_{2^{n-1}-1-2i}+\alpha _{4}d_{(2^{n}-1)-2i}, 
\notag \\
&&a_{2i+1}=\alpha _{1}d_{2i+1}+\alpha _{2}d_{2^{n-1}+2i+1},  \notag \\
&&a_{(2^{n}-2)-2i}=\alpha _{3}d_{(2^{n-1}-2)-2i}+\alpha _{4}d_{(2^{n}-2)-2i}.
\label{D3}
\end{eqnarray}

By (\ref{D3}), 
\begin{eqnarray}
&&(a_{2i}a_{(2^{n}-1)-2i}-a_{2i+1}a_{(2^{n}-2)-2i})=  \notag \\
&&\alpha _{1}\alpha _{3}(d_{2i}d_{2^{n-1}-1-2i}-d_{2i+1}d_{(2^{n-1}-2)-2i}) 
\notag \\
&&+\alpha _{1}\alpha _{4}(d_{2i}d_{(2^{n}-1)-2i}-d_{2i+1}d_{(2^{n}-2)-2i}) 
\notag \\
&&+\alpha _{2}\alpha
_{3}(d_{2^{n-1}+2i}d_{2^{n-1}-1-2i}-d_{2^{n-1}+2i+1}d_{(2^{n-1}-2)-2i}) 
\notag \\
&&+\alpha _{2}\alpha
_{4}(d_{2^{n-1}+2i}d_{(2^{n}-1)-2i}-d_{2^{n-1}+2i+1}d_{(2^{n}-2)-2i}).
\label{D4}
\end{eqnarray}

By (\ref{D0}), 
\begin{eqnarray}
&&a_{(2^{n-1}-2)-2i}=\alpha _{1}d_{(2^{n-1}-2)-2i}+\alpha _{2}d_{2^{n}-2-2i},
\notag \\
&&a_{(2^{n-1}+1)+2i}=\alpha _{3}d_{2i+1}+\alpha _{4}d_{2^{n-1}+1+2i},  \notag
\\
&&a_{(2^{n-1}-1)-2i}=\alpha _{1}d_{(2^{n-1}-1)-2i}+\alpha _{2}d_{2^{n}-1-2i},
\notag \\
&&a_{2^{n-1}+2i}=\alpha _{3}d_{2i}+\alpha _{4}d_{2^{n-1}+2i}.  \label{D5}
\end{eqnarray}%
So, by (\ref{D5}), 
\begin{eqnarray}
&&(a_{(2^{n-1}-2)-2i}a_{(2^{n-1}+1)+2i}-a_{(2^{n-1}-1)-2i}a_{2^{n-1}+2i})= 
\notag \\
&&-\alpha _{1}\alpha _{3}(d_{2i}d_{2^{n-1}-1-2i}-d_{2i+1}d_{(2^{n-1}-2)-2i})
\notag \\
&&+\alpha _{1}\alpha
_{4}(d_{(2^{n-1}-2)-2i}d_{2^{n-1}+1+2i}-d_{(2^{n-1}-1)-2i}d_{2^{n-1}+2i}) 
\notag \\
&&+\alpha _{2}\alpha _{3}(d_{2i+1}d_{2^{n}-2-2i}-d_{2i}d_{2^{n}-1-2i}) 
\notag \\
&&-\alpha _{2}\alpha
_{4}(d_{2^{n-1}+2i}d_{(2^{n}-1)-2i}-d_{2^{n-1}+2i+1}d_{(2^{n}-2)-2i}).
\label{D6}
\end{eqnarray}

So, by (\ref{D4}) and (\ref{D6}), 
\begin{eqnarray}
&&(a_{2i}a_{(2^{n}-1)-2i}-a_{2i+1}a_{(2^{n}-2)-2i})+(a_{(2^{n-1}-2)-2i}a_{(2^{n-1}+1)+2i}-a_{(2^{n-1}-1)-2i}a_{2^{n-1}+2i})=
\notag \\
&&\alpha _{1}\alpha
_{4}[(d_{2i}d_{(2^{n}-1)-2i}-d_{2i+1}d_{(2^{n}-2)-2i})+(d_{(2^{n-1}-2)-2i}d_{2^{n-1}+1+2i}-d_{(2^{n-1}-1)-2i}d_{2^{n-1}+2i})]
\notag \\
&&-\alpha _{2}\alpha
_{3}[(d_{2i}d_{(2^{n}-1)-2i}-d_{2i+1}d_{(2^{n}-2)-2i})+(d_{(2^{n-1}-2)-2i}d_{2^{n-1}+1+2i}-d_{(2^{n-1}-1)-2i}d_{2^{n-1}+2i})]=
\notag \\
&&[(d_{2i}d_{(2^{n}-1)-2i}-d_{2i+1}d_{(2^{n}-2)-2i})+(d_{(2^{n-1}-2)-2i}d_{(2^{n-1}+1)+2i}-d_{(2^{n-1}-1)-2i}d_{2^{n-1}+2i})]
\notag \\
&&\ast \det (\alpha ).  \notag
\end{eqnarray}

Lemma 2.

When $0\leq i\leq 2^{n-3}-1$, $sign^{\ast }(n-1,i)=sign(n,i)$.

Proof. There are two cases.

Case 1. $0\leq i\leq 2^{n-4}-1$.

By the definitions, $sign^{\ast }(n-1,i)=sign(n-1,i)$ and $%
sign(n,i)=sign(n-1,i)$. Therefore for the case, $sign^{\ast
}(n-1,i)=sign(n,i)$.

Case 2. $2^{n-4}-1<i\leq 2^{n-3}-1$.

By the definitions $sign^{\ast }(n-1,i)=sign(n-1,2^{n-3}-1-i)$ and $%
sign(n,i)=sign(n,2^{n-3}-1-i)$ because $n$ is odd. Since $0\leq
2^{n-3}-1-i<2^{n-4}$, by the definition $%
sign(n,2^{n-3}-1-i)=sign(n-1,2^{n-3}-1-i)$. Hence, $sign^{\ast
}(n-1,i)=sign(n,i)$ for the case.

Consequently, the argument is done by Cases 1 and 2.

\textbf{Part 1.} The proof of Theorem 1 (for even $n$-qubits)

For the proof of the invariant for 4-qubits, see Appendix A. The proof of
Theorem 1 follows the following Steps 1 and 2.

Step 1. Prove $IV(a,n)=IV(d,n)\det (\alpha )$, where $IV(d,n)$ is obtained
from $IV(a,n)$ by replacing $a$ by $d$.

By lemma 1 above, clearly Step 1 holds.

Step 2. Prove $IV(d,n)=IV(b,n)\underbrace{\det (\beta )\det (\gamma )...}%
_{n-1}$.

Step 2.1. Prove $IV(d,n)=IV(h,n)\det (\beta )$, where $%
\sum_{i=0}^{2^{n}-1}h_{i}|i\rangle =\underbrace{I\otimes I\otimes \gamma
.....}_{n}\sum_{i=0}^{2^{n}-1}b_{i}|i\rangle $ and $IV(h,n)$ is obtained
from $IV(a,n)$ by replacing $a$ by $h$.

Notice that in Step 2.1 we will present the idea which will be used in the
proof of Step 2.2 (for general case).

Proof.

From (\ref{D1}), 
\begin{eqnarray}
\sum_{i=0}^{2^{n-1}-1}d_{i}|i\rangle =(\beta _{1}|0\rangle +\beta
_{3}|1\rangle )\otimes \gamma \otimes
...\sum_{i=0}^{2^{n-2}-1}b_{i}|i\rangle +(\beta _{2}|0\rangle +\beta
_{4}|1\rangle )\otimes \gamma \otimes ...
\sum_{i=0}^{2^{n-2}-1}b_{2^{n-2}+i}|i\rangle.  \label{EE11}
\end{eqnarray}

Let 
\begin{eqnarray}
\sum_{i=0}^{2^{n-2}-1}h_{i}|i\rangle =\gamma \otimes
...\sum_{i=0}^{2^{n-2}-1}b_{i}|i\rangle \quad \mbox{ and} \quad
\sum_{i=0}^{2^{n-2}-1}h_{2^{n-2}+i}|i\rangle =\gamma \otimes ...
\sum_{i=0}^{2^{n-2}-1}b_{2^{n-2}+i}|i\rangle  \label{EX1}
\end{eqnarray}

Then (\ref{EE11}) can be rewritten as follows. 
\begin{equation*}
\sum_{i=0}^{2^{n-1}-1}d_{i}|i\rangle =|0\rangle \otimes
\sum_{i=0}^{2^{n-2}-1}(\beta _{1}h_{i}+\beta _{2}h_{2^{n-2}+i})|i\rangle
+|1\rangle \otimes \sum_{i=0}^{2^{n-2}-1}(\beta _{3}h_{i}+\beta
_{4}h_{2^{n-2}+i})|i\rangle.
\end{equation*}

Thus 
\begin{eqnarray}
d_{i}=\beta _{1}h_{i}+\beta _{2}h_{2^{n-2}+i} \quad \mbox{and}\quad
d_{2^{n-2}+i}=\beta _{3}h_{i}+\beta _{4}h_{2^{n-2}+i}, 0\leq i\leq 2^{n-2}-1.
\label{E3}
\end{eqnarray}

As well, from (\ref{D2}) we obtain 
\begin{eqnarray}
&&\sum_{i=0}^{2^{n-1}-1}d_{2^{n-1}+i}|i\rangle =  \notag \\
&&|0\rangle \otimes \sum_{i=0}^{2^{n-2}-1}(\beta _{1}h_{2^{n-1}+i}+\beta
_{2}h_{2^{n-1}+2^{n-2}+i})|i\rangle +|1\rangle \otimes
\sum_{i=0}^{2^{n-2}-1}(\beta _{3}h_{2^{n-1}+i}+\beta
_{4}h_{2^{n-1}+2^{n-2}+i})|i\rangle,  \label{E4}
\end{eqnarray}

where 
\begin{equation*}
\sum_{i=0}^{2^{n-2}-1}h_{2^{n-1}+i}|i\rangle =\gamma \otimes
...\sum_{i=0}^{2^{n-2}-1}b_{2^{n-1}+i}|i\rangle
\end{equation*}
and 
\begin{eqnarray}
\sum_{i=0}^{2^{n-2}-1}h_{2^{n-1}+2^{n-2}+i}|i\rangle =\gamma \otimes ...
\sum_{i=0}^{2^{n-2}-1}b_{2^{n-1}+2^{n-2}+i}|i\rangle  \label{EX2}
\end{eqnarray}

From (\ref{E4}), we obtain 
\begin{eqnarray}
d_{2^{n-1}+i}=\beta _{1}h_{2^{n-1}+i}+\beta _{2}h_{2^{n-1}+2^{n-2}+i}\quad%
\mbox{and}\quad d_{2^{n-1}+2^{n-2}+i}=\beta _{3}h_{2^{n-1}+i}+\beta
_{4}h_{2^{n-1}+2^{n-2}+i},  \label{E5}
\end{eqnarray}
where $0\leq i\leq 2^{n-2}-1$.

Note that from (\ref{EX1}) and (\ref{EX2}), clearly 
\begin{equation*}
\sum_{i=0}^{2^{n}-1}h_{i}|i\rangle =I\otimes I\otimes \gamma \otimes
...\sum_{i=0}^{2^{n}-1}b_{i}|i\rangle.
\end{equation*}

Now we demonstrate $IV(d,n)=IV(h,n)\det (\beta )$.

To compute $IV(d,n)$, let 
\begin{equation*}
T(i)=(d_{2i}d_{(2^{n}-1)-2i}-d_{2i+1}d_{(2^{n}-2)-2i})+(d_{(2^{n-1}-2)-2i}d_{(2^{n-1}+1)+2i}-d_{(2^{n-1}-1)-2i}d_{2^{n-1}+2i})
\end{equation*}
in (\ref{EvenIV1}).

Let us compute $T(i)$ by using (\ref{E3}) and (\ref{E5}). Then we obtain the
coefficients of $\beta _{1}\beta _{4},\beta _{2}\beta _{3},\beta _{1}\beta
_{3}$ and $\beta _{2}\beta _{4}$ in $T(i)$ as follows.

(1). The coefficients of $\beta _{1}\beta _{4}$ in $T(i)$ is

\begin{equation*}
sign(n,i)[(h_{2i}h_{(2^{n}-1)-2i}-h_{2i+1}h_{(2^{n}-2)-2i})+(h_{(2^{n-1}-2)-2i}h_{(2^{n-1}+1)+2i}-h_{(2^{n-1}-1)-2i}h_{2^{n-1}+2i})] .
\end{equation*}

Then it is easy to see that the coefficient of $\beta _{1}\beta _{4}$ in $%
IV(d,n)$ is $IV(h,n)$. \ \ 

(2). The coefficient of $\beta _{2}\beta _{3}$ in $T(i)$ is 
\begin{equation*}
sign(n,i)[(h_{2^{n-2}+2i}h_{3\ast 2^{n-2}-1-2i}-h_{2i+1+2^{n-2}}h_{3\ast
2^{n-2}-2-2i})+(h_{(2^{n-2}-2)-2i}h_{3\ast
2^{n-2}+1+2i}-h_{(2^{n-2}-1)-2i}h_{3\ast 2^{n-2}+2i})].
\end{equation*}

Then, the coefficient of $\beta _{2}\beta _{3}$ in $IV(d,n)$ is 
\begin{equation*}
\sum_{i=0}^{2^{n-3}-1}sign(n,i)[(h_{2^{n-2}+2i}h_{3\ast
2^{n-2}-1-2i}-h_{2i+1+2^{n-2}}h_{3\ast
2^{n-2}-2-2i})+(h_{(2^{n-2}-2)-2i}h_{3\ast
2^{n-2}+1+2i}-h_{(2^{n-2}-1)-2i}h_{3\ast 2^{n-2}+2i})].
\end{equation*}

Let $j=2^{n-3}-1-i$. Note that $sign(n,2^{n-3}-1-j)=-sign(n,j)$ by the
definition.\ It is not hard to see that the coefficient of $\beta _{2}\beta
_{3}$ in $IV(d,n)$ happens to be $-IV(h,n)$.

(3). The coefficient of $\beta _{1}\beta _{3}$ in $T(i)$ is 
\begin{equation*}
sign(n,i)[(h_{2i}h_{(3\ast 2^{n-2}-1)-2i}-h_{2i+1}h_{(3\ast
2^{n-2}-2)-2i})+(h_{(2^{n-2}-2)-2i}h_{(2^{n-1}+1)+2i}-h_{(2^{n-2}-1)-2i}h_{2^{n-1}+2i})] .
\end{equation*}

Note that the coefficient of $\beta _{1}\beta _{3}$ in $T(2^{n-3}-1-i)$ is
the opposite number of the one of $\beta _{1}\beta _{3}$ in $T(i)$ because $%
sign(n,2^{n-3}-1-i)=-sign(n,i)$. \ Therefore the coefficient of $\beta
_{1}\beta _{3}$ in $IV(d,n)$\ vanishes.

(4). The coefficient of $\beta _{2}\beta _{4}$ in $T(i)$ is

\begin{equation*}
sign(n,i)[(h_{2^{n-2}+2i}h_{(2^{n}-1)-2i}-h_{2^{n-2}+2i+1}h_{(2^{n}-2)-2i})+(h_{(2^{n-1}-2)-2i}h_{(3\ast 2^{n-2}+1)+2i}-h_{(2^{n-1}-1)-2i}h_{3\ast 2^{n-2}+2i})].
\end{equation*}

Note that the coefficient of $\beta _{2}\beta _{4}$ in $T(2^{n-3}-1-i)$ is
the opposite number of the one of $\beta _{2}\beta _{4}$ in $T(i)$. As well,
the coefficient of $\beta _{2}\beta _{4}$ in $IV(d,n)$\ vanishes.

From the above discussion, it is straightforward that $IV(d,n)=IV(h,n)\det
(\beta )$.

Step 2.2. For general case

Let 
\begin{equation*}
\sum_{i=0}^{2^{n}-1}p_{i}|i\rangle =\underbrace{I\otimes I\otimes ...\otimes
I}_{l}\otimes \underbrace{\tau \otimes \sigma \otimes ......}%
_{n-l}\sum_{i=0}^{2^{n}-1}b_{i}|i\rangle.
\end{equation*}

Then $IV(p,n)=IV(r,n)\det (\tau )$, where 
\begin{equation*}
\sum_{i=0}^{2^{n}-1}r_{i}|i\rangle =\underbrace{I\otimes I\otimes ...\otimes
I}_{l+1}\otimes \underbrace{\sigma \otimes ......}_{n-l-1}%
\sum_{i=0}^{2^{n}-1}b_{i}|i\rangle .
\end{equation*}
Note that $IV(p,n)$ and $IV(r,n)$ are obtained from $IV(a,n)$ by replacing $%
a $ by $p$ and $r$, respectively.

Proof.

We rewrite 
\begin{eqnarray}
&&\sum_{i=0}^{2^{n}-1}b_{i}|i\rangle =|0\rangle _{l}\otimes
\sum_{i=0}^{2^{n-l}-1}b_{i}|i\rangle _{n-l}+....+|k\rangle _{l}\otimes
\sum_{i=0}^{2^{n-l}-1}b_{k\ast 2^{n-l}+i}|i\rangle
_{n-l}+...+|2^{l}-1\rangle _{l}\otimes
\sum_{i=0}^{2^{n-l}-1}b_{(2^{l}-1)\ast 2^{n-l}+i}|i\rangle _{n-l}  \notag \\
&&=\sum_{k=0}^{2^{l}-1}(|k\rangle _{l}\otimes \sum_{i=0}^{2^{n-l}-1}b_{k\ast
2^{n-l}+i}|i\rangle _{n-l}).  \notag
\end{eqnarray}

Then 
\begin{equation*}
\sum_{i=0}^{2^{n}-1}p_{i}|i\rangle =\sum_{k=0}^{2^{l}-1}(|k\rangle
_{l}\otimes \underbrace{\tau \otimes \sigma \otimes ......}%
_{n-l}\sum_{i=0}^{2^{n-l}-1}b_{k\ast 2^{n-l}+i}|i\rangle _{n-l}).
\end{equation*}

Thus, $\sum_{i=0}^{2^{n-l}-1}p_{k\ast 2^{n-l}+i}|i\rangle =\underbrace{\tau
\otimes \sigma \otimes ......}_{n-l}\sum_{i=0}^{2^{n-l}-1}b_{k\ast
2^{n-l}+i}|i\rangle _{n-l}$, where $0\leq k\leq 2^{l}-1$.

By the above discussion, 
\begin{eqnarray}
&&\sum_{i=0}^{2^{n-l}-1}p_{k\ast 2^{n-l}+i}|i\rangle =(\tau _{1}|0\rangle
+\tau _{3}|1\rangle )\otimes \underbrace{\sigma \otimes ......}%
_{n-l-1}\sum_{i=0}^{2^{n-l-1}-1}b_{k\ast 2^{n-l}+i}|i\rangle _{n-l-1}  \notag
\\
&&+(\tau _{2}|0\rangle +\tau _{4}|1\rangle )\otimes \underbrace{\sigma
\otimes ......}_{n-l-1}\sum_{i=0}^{2^{n-l-1}-1}b_{k\ast
2^{n-l}+2^{n-l-1}+i}|i\rangle _{n-l-1}.  \label{E7}
\end{eqnarray}

Let 
\begin{eqnarray}
\sum_{i=0}^{2^{n-l-1}-1}r_{k\ast 2^{n-l}+i}|i\rangle =\underbrace{\sigma
\otimes ......}_{n-l-1}\sum_{i=0}^{2^{n-l-1}-1}b_{k\ast 2^{n-l}+i}|i\rangle
_{n-l-1}.  \label{E8}
\end{eqnarray}

and 
\begin{eqnarray}
\sum_{i=0}^{2^{n-l-1}-1}r_{k\ast 2^{n-l}+2^{n-l-1}+i}|i\rangle =\underbrace{%
\sigma \otimes ......}_{n-l-1}\sum_{i=0}^{2^{n-l-1}-1}b_{k\ast
2^{n-l}+2^{n-l-1}+i}|i\rangle _{n-l-1},  \label{E9}
\end{eqnarray}

where $0\leq k\leq 2^{l}-1$.

From (\ref{E8}) and (\ref{E9}), it is not hard to see that 
\begin{equation*}
\sum_{i=0}^{2^{n}-1}r_{i}|i\rangle =\underbrace{I\otimes ...\otimes I}%
_{l+1}\otimes \underbrace{\sigma \otimes ......}_{n-l-1}%
\sum_{i=0}^{2^{n}-1}b_{i}|i\rangle.
\end{equation*}

Then, from (\ref{E7}), (\ref{E8}) and (\ref{E9}) 
\begin{eqnarray}
&&\sum_{i=0}^{2^{n-l}-1}p_{k\ast 2^{n-l}+i}|i\rangle =|0\rangle \otimes
\sum_{i=0}^{2^{n-l-1}-1}(\tau _{1}r_{k\ast 2^{n-l}+i}+\tau _{2}r_{k\ast
2^{n-l}+2^{n-l-1}+i})|i\rangle _{n-l-1}  \notag \\
&&+|1\rangle \otimes \sum_{i=0}^{2^{n--l-1}-1}(\tau _{3}r_{k\ast
2^{n-l}+i}+\tau _{4}r_{k\ast 2^{n-l}+2^{n-l-1}+i})|i\rangle _{n-l-1}.
\label{E10}
\end{eqnarray}

Thus, from (\ref{E10}) 
\begin{eqnarray}
p_{k\ast 2^{n-l}+i}=\tau _{1}r_{k\ast 2^{n-l}+i}+\tau _{2}r_{k\ast
2^{n-l}+2^{n-l-1}+i}, p_{k\ast 2^{n-l}+2^{n-l-1}+i}=\tau _{3}r_{k\ast
2^{n-l}+i}+\tau _{4}r_{k\ast 2^{n-l}+2^{n-l-1}+i},  \label{E11}
\end{eqnarray}

where $0\leq k\leq 2^{l}-1$ and $0\leq i\leq 2^{n-l-1}-1$.\ 

By using the idea used in Step 2.1 above, \ from (\ref{E11})\ we can show $%
IV(p,n)=IV(r,n)\det (\tau )$.

Conclusively, it is not hard to prove Step 2 by repeating applications of
Step 2.2.

\textbf{Part 2}. The proof of Theorem 2 (for odd $n$-qubits)

For the proofs for $3$-qubits and $5$-qubits, see Appendixes B and C,
respectively.







The proof of Theorem 2 follows the following Steps 1 and 2 immediately.

Step 1. Prove 
\begin{equation*}
(\overline{IV}(a,n))^{2}-4IV^{\ast }(a,n-1)IV_{+2^{n-1}}^{\ast }(a,n-1)=(%
\overline{IV}(d,n))^{2}-4IV^{\ast }(d,n-1)IV_{+2^{n-1}}^{\ast
}(d,n-1)\det^{2}(\alpha ),
\end{equation*}

where $\overline{IV}(d,n)$, $IV^{\ast }(d,n-1)$ and $IV_{+2^{n-1}}^{\ast
}(d,n-1)$ are obtained from $\overline{IV}(a,n)$, $IV^{\ast }(a,n-1)$ and $%
IV_{+2^{n-1}}^{\ast }(a,n-1)$ by replacing $a$ by $d$, respectively.

Step 1.1. Prove 
\begin{equation*}
IV^{\ast }(a,n-1)=IV^{\ast }(d,n-1)\alpha _{1}^{2}+ \overline{IV}(d,n)\alpha
_{1}\alpha _{2}+IV_{+2^{n-1}}^{\ast }(d,n-1)\alpha _{2}^{2}.
\end{equation*}

By the definition, 
\begin{equation*}
IV^{\ast }(a,n-1)=\sum_{i=0}^{2^{n-3}-1}sign^{\ast
}(n-1,i)(a_{2i}a_{(2^{n-1}-1)-2i}-a_{2i+1}a_{(2^{n-1}-2)-2i}).
\end{equation*}

When $0\leq i\leq 2^{n-3}-1$, clearly 
\begin{equation*}
0\leq 2i,2i+1,(2^{n-1}-2)-2i,(2^{n-1}-1)-2i\leq (2^{n-1}-1).
\end{equation*}

Hence, from (\ref{D0}), 
\begin{eqnarray}
&&a_{2i}=\alpha _{1}d_{2i}+\alpha _{2}d_{2^{n-1}+2i},\quad
a_{(2^{n-1}-1)-2i}=\alpha _{1}d_{(2^{n-1}-1)-2i}+\alpha _{2}d_{2^{n}-1-2i}, 
\notag \\
&&a_{2i+1}=\alpha _{1}d_{2i+1}+\alpha _{2}d_{2^{n-1}+2i+1},\quad
a_{(2^{n-1}-2)-2i}=\alpha _{1}d_{(2^{n-1}-2)-2i}+\alpha _{2}d_{2^{n}-2-2i}.
\label{amp1}
\end{eqnarray}

By substituting (\ref{amp1}) into $IV^{\ast }(a,n-1)$, 
\begin{eqnarray}
&&IV^{\ast }(a,n-1)=  \notag \\
&&\alpha _{1}^{2}\sum_{i=0}^{2^{n-3}-1}sign^{\ast
}(n-1,i)(d_{2i}d_{(2^{n-1}-1)-2i}-d_{2i+1}d_{(2^{n-1}-2)-2i})  \notag \\
&&+\alpha _{1}\alpha _{2}\sum_{i=0}^{2^{n-3}-1}sign^{\ast
}(n-1,i)[(d_{2i}d_{(2^{n}-1)-2i}-d_{2i+1}d_{(2^{n}-2)-2i})-(d_{(2^{n-1}-2)-2i}d_{(2^{n-1}+1)+2i}-d_{(2^{n-1}-1)-2i}d_{2^{n-1}+2i})]
\notag \\
&&+\alpha _{2}^{2}\sum_{i=0}^{2^{n-3}-1}sign^{\ast
}(n-1,i)(d_{2^{n-1}+2i}d_{(2^{n}-1)-2i}-d_{2^{n-1}+2i+1}d_{(2^{n}-2)-2i}) 
\notag \\
&=&IV^{\ast }(d,n-1)\alpha _{1}^{2}+\overline{IV}(d,n)\alpha _{1}\alpha
_{2}+IV_{+2^{n-1}}^{\ast }(d,n-1)\alpha _{2}^{2}.  \label{F3}
\end{eqnarray}

Step 1.2. Calculating $IV_{+2^{n-1}}^{\ast }(a,n-1)$

As discussed in Step 1.1, we can demonstrate 
\begin{equation*}
IV_{+2^{n-1}}^{\ast }(a,n-1)=IV^{\ast }(d,n-1)\alpha _{3}^{2}+\overline{IV}%
(d,n)\alpha _{3}\alpha _{4}+IV_{+2^{n-1}}^{\ast }(d,n-1)\alpha _{4}^{2}.
\end{equation*}

Step 1.3. Prove 
\begin{equation*}
\overline{IV}(a,n)=2\ast IV^{\ast }(d,n-1)\alpha _{1}\alpha _{3}+\overline{IV%
}(d,n)(\alpha _{1}\alpha _{4}+\alpha _{2}\alpha _{3})+2\ast
IV_{+2^{n-1}}^{\ast }(d,n-1)\alpha _{2}\alpha _{4}.
\end{equation*}

By the definition, 
\begin{equation*}
\overline{IV}(a,n)=%
\sum_{i=0}^{2^{n-3}-1}sign(n,i)[(a_{2i}a_{(2^{n}-1)-2i}-a_{2i+1}a_{(2^{n}-2)-2i})-(a_{(2^{n-1}-2)-2i}a_{(2^{n-1}+1)+2i}-a_{(2^{n-1}-1)-2i}a_{2^{n-1}+2i})] .
\end{equation*}

When $0\leq i\leq 2^{n-3}-1$, clearly 
\begin{equation*}
2^{n-1}-1<(2^{n}-1)-2i,(2^{n}-2)-2i,(2^{n-1}+1)+2i,2^{n-1}+2i.
\end{equation*}

Therefore, by (\ref{D0}) 
\begin{eqnarray}
&&a_{(2^{n}-1)-2i}=\alpha _{3}d_{(2^{n-1}-1)-2i}+\alpha
_{4}d_{2^{n}-1-2i},a_{(2^{n}-2)-2i}=\alpha _{3}d_{(2^{n-1}-2)-2i}+\alpha
_{4}d_{2^{n}-2-2i},  \notag \\
&&a_{(2^{n-1}+1)+2i}=\alpha _{3}d_{2i+1}+\alpha
_{4}d_{2^{n-1}+1+2i},a_{2^{n-1}+2i}=\alpha _{3}d_{2i}+\alpha
_{4}d_{2^{n-1}+2i}.  \label{amp2}
\end{eqnarray}

By substituting (\ref{amp1}) and (\ref{amp2}) and computing, 
\begin{eqnarray}
&&(a_{2i}a_{(2^{n}-1)-2i}-a_{2i+1}a_{(2^{n}-2)-2i})-(a_{(2^{n-1}-2)-2i}a_{(2^{n-1}+1)+2i}-a_{(2^{n-1}-1)-2i}a_{2^{n-1}+2i})=
\notag \\
&&2(d_{2i}d_{(2^{n-1}-1)-2i}-d_{2i+1}d_{(2^{n-1}-2)-2i})\alpha _{1}\alpha
_{3}  \notag \\
&&+[(d_{2i}d_{(2^{n}-1)-2i}-d_{2i+1}d_{(2^{n}-2)-2i})-(d_{(2^{n-1}-2)-2i}d_{2^{n-1}+2i+1}-d_{(2^{n-1}-1)-2i}d_{2^{n-1}+2i})](\alpha _{1}\alpha _{4}+\alpha _{2}\alpha _{3})
\notag \\
&&+2(d_{2^{n-1}+2i}d_{(2^{n}-1)-2i}-d_{2^{n-1}+2i+1}d_{(2^{n}-2)-2i})\alpha
_{2}\alpha _{4}.  \notag
\end{eqnarray}

Note that when $0\leq i\leq 2^{n-3}-1$, $sign^{\ast }(n,i)=sign(n,i)$ by the
definition and $sign(n,i)=sign^{\ast }(n-1,i)$ by lemma 2. Thus, the proof
of Step 1.3 is done.

By Steps 1.1, 1.2 and 1.3, we finish the proof of Step 1.

Step 2. Prove that 
\begin{eqnarray}
&&(\overline{IV}(d,n))^{2}-4IV^{\ast }(d,n-1)IV_{+2^{n-1}}^{\ast }(d,n-1)= 
\notag \\
&& [(\overline{IV}(b,n))^{2}-4IV^{\ast }(b,n-1)IV_{+2^{n-1}}^{\ast }(b,n-1)]%
\underbrace{\det^{2}(\beta )\det^{2}(\gamma )...}_{n-1}.  \notag
\end{eqnarray}

By Theorem 1 for $(n-1)$-qubits, from (\ref{D1}), 
\begin{equation}
IV^{\ast }(d,n-1)=IV^{\ast }(b,n-1)\underbrace{\det (\beta )\det (\gamma )...%
}_{n-1}  \label{F5}
\end{equation}

and from (\ref{D2}) 
\begin{eqnarray}
IV_{+2^{n-1}}^{\ast }(d,n-1)=IV_{+2^{n-1}}^{\ast }(b,n-1) \underbrace{\det
(\beta )\det (\gamma )...}_{n-1}.  \label{F6}
\end{eqnarray}

Let us compute $\overline{IV}(d,n)$. From (\ref{D1}) and (\ref{D2}) we
obtain 
\begin{eqnarray}
\sum_{i=0}^{2^{n-1}}(d_{i}-d_{2^{n-1}+i})|i\rangle =\underbrace{\beta
\otimes \gamma ...}_{n-1}\sum_{i=0}^{2^{n-1}}(b_{i}-b_{2^{n-1}+i})|i\rangle .
\label{F7}
\end{eqnarray}

Let $d_{i}^{\ast }=d_{i}-d_{2^{n-1}+i}$ and $b_{i}^{\ast
}=b_{i}-b_{2^{n-1}+i}$. Then (\ref{F7}) can be rewritten as 
\begin{eqnarray}
\sum_{i=0}^{2^{n-1}}d_{i}^{\ast }|i\rangle =\underbrace{\beta \otimes \gamma
...}_{n-1}\sum_{i=0}^{2^{n-1}}b_{i}^{\ast }|i\rangle .  \label{F8}
\end{eqnarray}

By Theorem 1 for $(n-1)$-qubits, from (\ref{F8}) it is easy to see 
\begin{equation}
IV^{\ast }(d^{\ast },n-1)=IV^{\ast }(b^{\ast },n-1)\underbrace{\det (\beta
)\det (\gamma )...}_{n-1}.)  \label{F9}
\end{equation}

Note that 
\begin{equation*}
IV^{\ast }(d^{\ast },n-1)=\sum_{i=0}^{2^{n-3}-1}sign^{\ast
}(n-1,i)(d_{2i}^{\ast }d_{(2^{n-1}-1)-2i}^{\ast }-d_{2i+1}^{\ast
}d_{(2^{n-1}-2)-2i}^{\ast })
\end{equation*}%
and $sign^{\ast }(n-1,i)=sign(n,i)$ whenever $0\leq i\leq 2^{n-3}-1$ by (\ref%
{F4}).

By expanding, 
\begin{eqnarray}
IV^{\ast }(d^{\ast },n-1)=IV^{\ast }(d,n-1)+IV_{+2^{n-1}}^{\ast }(d,n-1)-%
\overline{IV}(d,n). )  \label{F10}
\end{eqnarray}

Similarly, by expanding, 
\begin{eqnarray}
IV(b^{\ast },n-1)=IV^{\ast }(b,n-1)+IV_{+2^{n-1}}^{\ast }(b,n-1)-\overline{IV%
}(b,n).  \label{F11}
\end{eqnarray}

Thus, substituting (\ref{F10}) and (\ref{F11}) into (\ref{F9}), we have 
\begin{eqnarray}
&&IV^{\ast }(d,n-1)+IV_{+2^{n-1}}^{\ast }(d,n-1)-\overline{IV}(d,n)=  \notag
\\
&&[IV^{\ast }(b,n-1)+IV_{+2^{n-1}}^{\ast }(b,n-1)-\overline{IV}(b,n)]%
\underbrace{\det (\beta )\det (\gamma )...}_{n-1}.  \label{F12}
\end{eqnarray}

From (\ref{F5}), (\ref{F6}) and (\ref{F12}), we get 
\begin{eqnarray}
\overline{IV}(d,n)=\overline{IV}(b,n) \underbrace{\det (\beta )\det (\gamma
)...}_{n-1}.  \label{F13}
\end{eqnarray}

The proof of Step 2 follows (\ref{F5}), (\ref{F6}) and (\ref{F13})
immediately.

\section*{Appendix E: The proof of $\protect\tau \leq 1$}

\setcounter{equation}{0} \renewcommand{\theequation}{E\arabic{equation}}

Let $f=(\overline{IV}(a,n))^{2}-4IV^{\ast }(a,n-1)IV_{+2^{n-1}}^{\ast
}(a,n-1)$ and $a_{i}$ be real. To find the extremes of $f$, we compute the
following partial derivatives:

\begin{equation}
\mbox{from}\quad \partial f/\partial a_{0}= 0, \overline{IV}
(a,n)sign(n,0)a_{2^{n}-1}=2IV_{+2^{n-1}}^{\ast }(a,n-1)sign^{\ast
}(n-1,0)a_{2^{n-1}-1};  \label{ext1}
\end{equation}

\begin{equation}
\mbox{from}\quad \partial f/\partial a_{1}= 0, \overline{IV}
(a,n)sign(n,0)a_{2^{n}-2}=2IV_{+2^{n-1}}^{\ast }(a,n-1)sign^{\ast
}(n-1,0)a_{2^{n-1}-2};  \label{ext2}
\end{equation}

........... 
\begin{equation}
\mbox{from}\quad \partial f/\partial a_{2^{n-1}-2}=0, \overline{IV}
(a,n)sign(n,0)a_{2^{n-1}+1}=2IV_{+2^{n-1}}^{\ast }(a,n-1)sign^{\ast
}(n-1,0)a_{1};  \label{ext3}
\end{equation}

\begin{equation}
\mbox{from}\quad \partial f/\partial a_{2^{n-1}-1}=0, \overline{IV}
(a,n)sign(n,0)a_{2^{n-1}}=2IV_{+2^{n-1}}^{\ast }(a,n-1)sign^{\ast
}(n-1,0)a_{0};  \label{ext4}
\end{equation}
\begin{equation}
\mbox{from}\quad \partial f/\partial a_{2^{n-1}}= 0, \overline{IV}
(a,n)sign(n,0)a_{2^{n-1}-1}=2IV^{\ast }(a,n-1)sign^{\ast }(n-1,0)a_{2^{n}-1}
;  \label{ext5}
\end{equation}
.......... 
\begin{equation}
\mbox{from}\quad \partial f/\partial a_{2^{n}-1}=0, \overline{IV}
(a,n)sign(n,0)a_{0}=2IV^{\ast }(a,n-1)sign^{\ast }(n-1,0)a_{2^{n-1}}.
\label{ext6}
\end{equation}

From (\ref{ext1}) $\times$ (\ref{ext4}), 
\begin{equation}
(\overline{IV} (a,n))^{2}sign(n,0)a_{2^{n-1}}a_{2^{n}-1}=4(IV_{+2^{n-1}}^{%
\ast }(a,n-1))^{2}sign^{\ast }(n-1,0)a_{0}a_{2^{n-1}-1}.  \label{ext7}
\end{equation}

From (\ref{ext2})$\times$ (\ref{ext3}), 
\begin{equation}
(\overline{IV} (a,n))^{2}sign(n,0)a_{2^{n-1}+1}a_{2^{n}-2}=4(IV_{+2^{n-1}}^{%
\ast }(a,n-1))^{2}sign^{\ast }(n-1,0)a_{1}a_{2^{n-1}-2}.  \label{ext8}
\end{equation}
........

From (\ref{ext7})$-$(\ref{ext8}), 
\begin{eqnarray}
&&(\overline{IV}
(a,n))^{2}sign(n,0)(a_{2^{n-1}}a_{2^{n}-1}-a_{2^{n-1}+1}a_{2^{n}-2})=  \notag
\\
&& 4(IV_{+2^{n-1}}^{\ast }(a,n-1))^{2}sign^{\ast
}(n-1,0)(a_{0}a_{2^{n-1}-1}-a_{1}a_{2^{n-1}-2}).  \label{ext9}
\end{eqnarray}
........

Evaluate the sum over the above expressions like (\ref{ext9}), we obtain 
\begin{equation}
(\overline{IV}(a,n))^{2}IV_{+2^{n-1}}^{\ast }(a,n-1)=4(IV_{+2^{n-1}}^{\ast
}(a,n-1))^{2}IV^{\ast }(a,n-1).  \label{ext10}
\end{equation}
As well, we have 
\begin{equation}
(\overline{IV}(a,n))^{2}IV^{\ast }(a,n-1)=4(IV^{\ast
}(a,n-1))^{2}IV_{+2^{n-1}}^{\ast }(a,n-1).  \label{ext11}
\end{equation}

From (\ref{ext10}), $IV_{+2^{n-1}}^{\ast }(a,n-1)=0$ or $f=0$. From (\ref%
{ext11}) , $IV^{\ast }(a,n-1)=0$ or $f=0$. When $IV_{+2^{n-1}}^{\ast
}(a,n-1)=0$ or $IV^{\ast }(a,n-1)=0$, it is not hard to see that $f=(%
\overline{IV}(a,n))^{2}\leq 1/4$. When $f=1/4$, $\left\vert a_{j}\right\vert
=\left\vert a_{2^{n}-1-j}\right\vert $.

Therefore $0\leq f\leq 1/4$ and $0\leq \tau \leq 1$.


\begin{thebibliography}{99}
\bibitem{Bennett} C. H. Bennett et al, quant-ph/9908073.

\bibitem{Bennett2} C. H. Bennett et al, Phys. Rev. A 63, 012307{2001}.

\bibitem{Acin} A. A\'{c}in et al., quant-ph/0003050.

\bibitem{Dur} W. D$\ddot{u}$r, G.Vidal and J.I. Cirac, Phys. Rev. A. 62
(2000)062314.

\bibitem{Acin1} A. Acin, E. Jane, W.D$\ddot{u}$r and G.vidal, Phys. Rev.
Lett. 85, 4811 (2000).

\bibitem{Briegel} H.J.Briegel and R.Raussendorf, Phy. Rev. Lett. 86, 910
(2001).

\bibitem{Lo} H.K.LO and S. Popescu, Phys. Rev. A. 63 02230 (2001).

\bibitem{Moor} F. Verstraete, J.Dehaene and B.De Moor, Phys. Rev. A. 65,
032308 (2002).

\bibitem{Moor2} F. Verstraete, J.Dehaene, B.De Moor and H. Verschelde Phys.
Rev. A. 65, 052112 (2002).

\bibitem{Miyake03} A. Miyake, PRA 67, 012108 (2003).

\bibitem{Miyake} A.Miyake, quant-ph/0401023.

\bibitem{Rajagopal} A.K. Rajagopal and R.W. Rendell, PRA 65, 032328 (2002).

\bibitem{LDF-PLA} D. Li et al., Simple criteria for the SLOCC
classification, Phys. Lett. A 359, 428(2006).

\bibitem{LDFJPA} D. Li et al., submitted to JPA, quant-ph/070132.

\bibitem{LDF} D. Li et al., the necessary and sufficient conditions for
separability for multipartite pure states, unpublished, submitted to PRL,
the paper No. LV9637(Sep. 2004) and quant-ph/0604147.

\bibitem{Coffman} V. Coffman et al., PRA 61, 052306 (2000).

\bibitem{Yu} Chang-shui Yu and He-shan Song, PRA 71, 042331(2005).

\bibitem{Wong} A. Wong and N. Christensen, PRA 63, 044301(2001).

\bibitem{Osterloh1} A. Osterloh and J. Siewert, PRA 72, 012337(2005).

\bibitem{Osterloh2} A. Osterloh and J. Siewert, International journal of
quantum information Vol. 4, No.3 (2006) 531-540.
\end{thebibliography}
\end{document}